\documentclass[lettersize,journal]{IEEEtran}
\usepackage{amsmath,amsfonts}
\usepackage{algorithmic}
\usepackage{algorithm}
\usepackage{array}
\usepackage[caption=false,font=normalsize,labelfont=sf,textfont=sf]{subfig}
\usepackage{textcomp}
\usepackage{stfloats}
\usepackage{url}
\usepackage{verbatim}
\usepackage{graphicx}
\usepackage{cite}
\usepackage{color}

{\begin{list}{}{\settowidth{\labelwidth}{\bf #1}%
			\setlength{\leftmargin}{\labelwidth}%
			\addtolength{\leftmargin}{\labelsep}%
			}}%
{\end{list}}

\begin{document}

\title{Gait-Based Privacy Protection for  Smart Wearable Devices}

\author{Yu Su, Yongjiao Li, Zhu Cao$^*$
\thanks{This work was supported in part by the National Natural Science Foundation of China (Key Program: 62136003), the National Natural Science Foundation of China (12105105), the Natural Science Foundation of Shanghai (21ZR1415800), the Shanghai Sailing Program (21YF1409800), and the startup fund from East China University of Science and Technology (YH0142214).}
\thanks{Y. Su, Y. Li, and Z. Cao are affiliated with Key Laboratory of Smart Manufacturing in Energy Chemical Process, Ministry of Education, East China University of Science and Technology, Shanghai. }
\thanks{*Corresponding author. Email address: caozhu@ecust.edu.cn (Zhu Cao).}
\thanks{Copyright (c) 20xx IEEE. Personal use of this material is permitted. However, permission to use this material for any other purposes must be obtained from the IEEE by sending a request to pubs-permissions@ieee.org.}

}

\maketitle

\begin{abstract}
Smart wearable devices (SWDs) collect and store sensitive daily information of many people. Its primary method of identification is still the password unlocking method. However, several studies have shown serious security flaws in that method, which makes the privacy and security concerns of SWDs particularly urgent. Gait identification is well suited for SWDs because its built-in sensors can provide data support for identification. However, existing gait identification methods have low accuracy and neglect to protect the privacy of gait features. In addition, the SWD can be used as an internet of things device for users to share data. But few studies have used gait feature-based encryption schemes to protect the privacy of message interactions between SWDs and other devices. In this paper, we propose a gait identification network, a bi-directional long short-term memory network with an attention mechanism (ABLSTM), to improve the identification accuracy and a stochastic orthogonal transformation (SOT) scheme to protect the extracted gait features from leakage. In the experiments, ABLSTM achieves an accuracy of 95.28$\%$, reducing previous error rate by 19.3$\%$. The SOT scheme is proved to be resistant to the chosen plaintext attack (CPA) and is 30$\%$ faster than previous methods. A biometric-based encryption scheme is proposed to enable secure message interactions using gait features as keys after the gait identification stage is passed, and offers better protection of the gait features compared to previous schemes.
\end{abstract}


\section{Introduction}
\IEEEPARstart{I}{n} recent years, smart wearable devices (SWDs) have brought great convenience to our living, entertainment, health, and payment~\cite{wang2016security,stephenson2022sok}. SWDs have a variety of built-in sensors that can collect and store people's daily sensitive information~\cite{bujari2018smart}. According to the International Data Corporation~\cite{networks}, global shipments of wearable devices grew by 9.9$\%$ in the third quarter of 2021, reaching 138.4 million units. SWDs are even used in the military, so their privacy and security issues have received much concern. Therefore, the SWD needs an identification mechanism to protect privacy.

Nowadays, identification schemes of the SWD are divided into the password unlocking method and the biometric identification method. The password unlocking method includes personal identification number (PIN) code unlocking~\cite{wang2017personal} and user-defined password combination~\cite{li2020designing,li2022secure}. The security risk in the SWD has been found in the password unlocking method~\cite{wang2017personal,li2020designing,li2022secure,ali2012zero,wang2016friend}. Guo et al.~\cite{wang2016security} used the built-in sensors of the SWD to identify the hand motion during password input and obtained the user's PIN code by distance estimation and backward derivation. Wang et al.~\cite{wang2016friend} implemented millimeter-level gesture determinations. In their experiments, they have a 90$\%$ probability of stealing the PIN code through gesture determination by three attempts to enter the PIN code.

Common biometric identification methods include fingerprints~\cite{aman2020lightweight}, face~\cite{kumar2021efficient}, iris~\cite{gad2019iot}, finger veins~\cite{das2018convolutional}, palmprints~\cite{genovese2019palmnet}, and voice~\cite{murphy2021integrating}. Biometric identification is considered an effective method for individual identification due to its uniqueness, stability, irreproducibility, and collectability~\cite{huang2009novel,1262027}. Users do not need to remember complex passwords or carry things like keys or smart cards. In addition, biometric identification is easy to work with the software for automated management. 

Among those biometric identifications, inertial measurement unit (IMU)-based gait identification is particularly suitable for SWDs. There are two reasons: First, the SWD has a built-in IMU sensor. The IMU can acquire three-axis acceleration and gyroscope signals~\cite{ilewicz2018estimation} to provide data support for gait identification.
Other biological identifications, such as face or fingerprint identification, are not suitable for the SWD, because the relevant hardware support is missing.
 Second, the gait is a complex spatiotemporal biological feature that is difficult to imitate and replicate~\cite{1262027}.

Among different gait identification methods \cite{ailisto2005identifying,rong2007identification,batchuluun2018gait,zhu2020one,tran2021multi,li2021novel,rua2021security,pan2023toward}, currently gait networks have the highest recognition accuracy. However, existing gait networks~\cite{batchuluun2018gait,zhu2020one,tran2021multi,li2021novel,rua2021security,pan2023toward} still do not have sufficiently high identification accuracy for practical deployment.  More importantly, the storage security of the feature templates for biometric identification is often overlooked, which can lead to the leakage and misuse of feature data. That would pose a security threat to future identification. In addition, the SWD can be used as an internet of things device for users to share data. But few studies have considered gait feature-based encryption schemes that protect the privacy of message interactions between SWDs and other devices.

In response to the above problems, this paper proposes a gait-based module capable of both identification and message interactions, called AmSoBe. AmSoBe consists of three novel parts: a bi-directional long and short-term memory with an attention network (ABLSTM), a stochastic orthogonal transformation (SOT) algorithm, and a biometric-based encryption (BBE) scheme. ABLSTM is used to deal with poor gait recognition accuracy in previous gait networks. SOT is used for the secure storage and identification of extracted gait features. BBE is used to enable secure message interactions. These three components of AmSoBe are connected as follows. 
First, ABLSTM extracts gait features from the raw gait signals. Second, SOT encrypts the gait features that are extracted from ABLSTM to protect the privacy of gait features. 
SOT also performs identification by comparing the encrypted gait features between the registration stage and the identification stage.
Finally, BBE uses the encrypted gait feature generated by SOT as keys to enable secure message interactions between SWD and other devices. Our main contributions are as follows:

$\bullet$ The ABLSTM network is proposed. Compared with a bi-directional long short-term memory (BLSTM) network, it can extract high-precision gait features and increase the reliability of identification. In the experiment, the identification accuracy is up to 95.28$\%$. Compared with the gait network in~\cite{tran2021multi}, the error rate is reduced by 19.3$\%$.

$\bullet$ To protect the privacy of gait features and prevent them from illegal use, the SOT scheme is designed, which is based on the Euclidean distance. It encrypts gait features using the orthogonal transformation and is proven to be resistant to chosen-plaintext attack (CPA). In experiments, it is tested against biometric templates of various lengths and with various numbers of randomly positioned parameters. The result shows that it encrypts 30$\%$ faster than other methods.

$\bullet$ Further, to help users easily encrypt messages using gait features and decrypt them after the gait features pass authentication, a BBE scheme is proposed which is based on the Euclidean distance and identity-based encryption (IBE) \cite{shamir1984identity}. Unlike fuzzy identity-based encryption (FIBE)~\cite{mao2016fully}, the biological template distance is used in the BBE to determine the similarity of biometric features. Finally, the BBE scheme is proved to be secure under the Decision Bilinear Diffie-Hellman (DBDH) assumption.

\section{Related Work}
\label{sec:rw}
\subsection{Gait Recognition}
Since the IMU sensor is small and portable, it can be placed anywhere on the body (wrist, waist, leg, shoe)~\cite{ngo2014largest,zou2020deep,zhang2014accelerometer,alsheikh2016deep}. This facilitates the collection of gait data and lays the foundation for IMU-based gait recognition. In early studies, Ailisto et al.~\cite{ailisto2005identifying} proposed to normalize and average the gait signals to equal amplitude and length. They performed gait recognition on 36 test subjects, and as a result, their scheme achieved a similar error rate of 6.4$\%$. The subsequent developments revolve around the dynamic time warping (DTW) algorithm and its improvements. Rong et al.~\cite{rong2007identification} found the gait cycle affects recognition and proposed the DTW algorithm. They recognized 35 test subjects with an error rate of 6.7$\%$. Chen et al.~\cite{chen2020gait} proposed a DTW-based $k$-nearest neighbor (kNN) algorithm to determine the beginning and end of the gait cycle. Their scheme combined with the template method achieved a micro-f1 score of 0.93. Pan et al.~\cite{pan2009accelerometer} proposed a voting algorithm based on signature points (SPs). It compared the extreme value points between gait templates for recognition, which avoids the problem of gait cycle misalignment caused by DTW. The algorithm extracts SPs that enable gait features to focus on local patterns of individual features. In recognition experiments with 30 subjects using five acceleration sensors, the recognition accuracy can reach 96.7$\%$. Since classifiers with SP clustering can better solve the gait cycle detection problem, Zhang et al.~\cite{zhang2014accelerometer} took the SP-based algorithm and proposed the sparse-code collection classifier. The error rate for validation is 2.2$\%$ on their published ZJU-GaitAcc dataset.

%---------------------------------------------------
\begin{figure*}[htb]
    \centering
    \includegraphics[width=18cm]{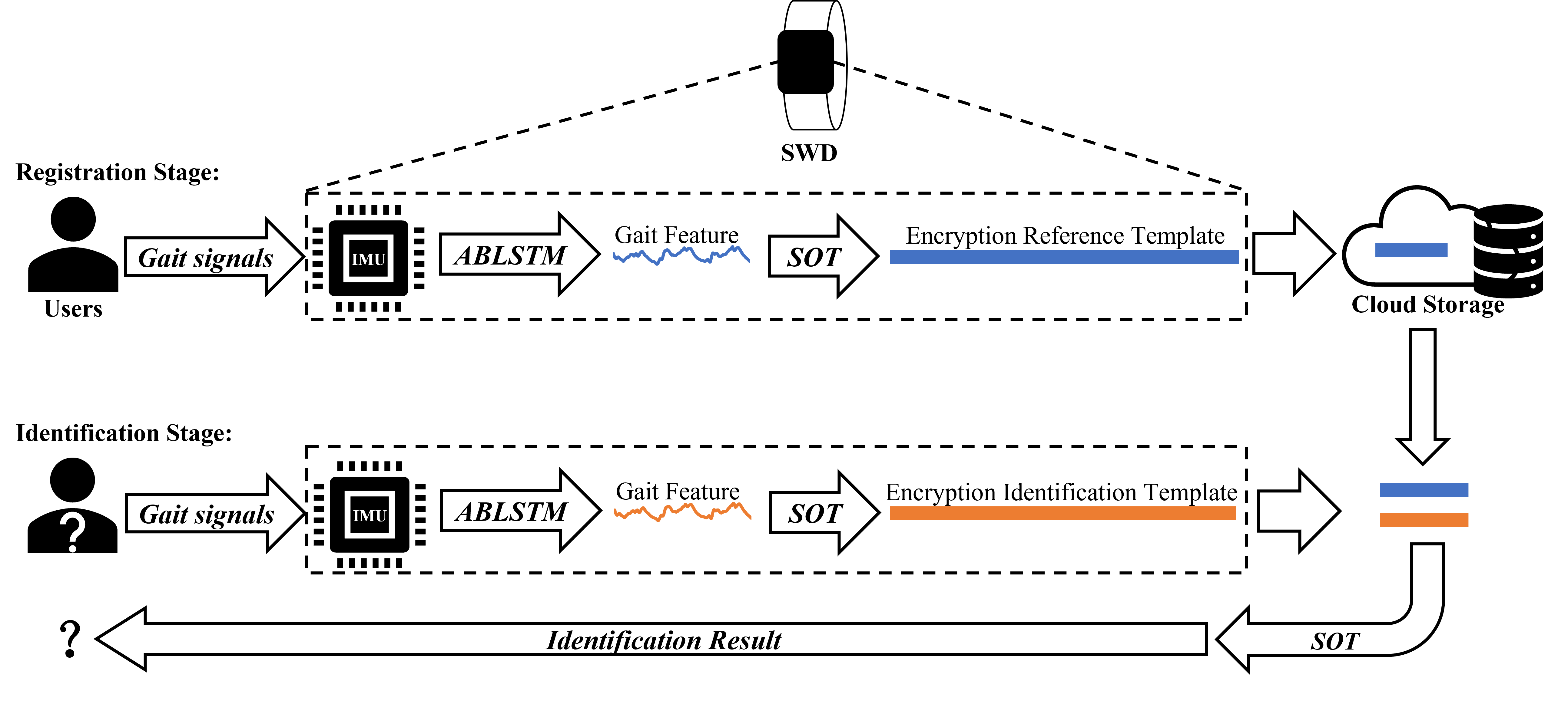}%18cm
    \caption{The process of the SWD gait registration stage and identification stage.}
    \label{sbt}
\end{figure*}
%% %-------------------------------------------------

The latest research widely uses deep learning methods for recognition, such as convolutional neural network (CNN)~\cite{8957063,asuncion2018thigh}, long short-term memory (LSTM)~\cite{tran2021multi,batchuluun2018gait} and CNN-LSTM hybrid models~\cite{li2021novel} to automatically extract gait features. Because it is based on gradient descent and updates the parameters of the network for each iteration, the network can achieve a high recognition accuracy after multiple rounds of training and validation. And the network can also be built with different architectures for other recognition purposes. Zou et al.~\cite{zou2020deep} proposed a CNN and LSTM hybrid network that can perform recognition in unconstrained or accessible environments. Their scheme combines traditional feature extraction and deep learning to improve recognition accuracy, demonstrating the effectiveness and accuracy of the deep learning networks.

\subsection{Privacy-Preserving Biometrics}
The biometric feature is innate and unique to everybody. If the feature is leaked, biometrics identification will face permanent security risks. However, there is no research on the privacy protection of gait templates. In contrast, privacy protection has been used for other biometric identifications. In the field of palmprint identification, Bouraoui et al.~\cite{bouraoui2015performance} proposed an error correction scheme with low correlation and high security. The experiment shows their identification system can extract features from the encrypted palmprint data for identification. After that, Lakhera et al.~\cite{lakhera2016efficient} used the advanced encryption standard (AES) algorithm and proved it is secure for scrambling the biometric template. Recently, Kolberg et al.~\cite{kolberg2020efficiency} used homomorphic encryption to resist attacks in the post-quantum era. Its advantage is relevant calculations can be directly performed with ciphertexts, but it is currently challenging to be implemented.

\subsection{Biometric-Based Encryption}
The proposed BBE scheme in this paper is based on identity-based encryption (IBE). The idea of IBE was first proposed in 1984~\cite{shamir1984identity}. In 2001, Boneh et al.~\cite{boneh2001identity} applied a bilinear map to construct a secure IBE scheme, thus formally defining the IBE security model. In 2004, Boneh and Boyen~\cite{boneh2004efficient} proposed a provably secure IBE method under a selective security attack model. Since then, there have been many advances~\cite{waters2005efficient,gentry2006practical,waters2009dual} in the security and effectiveness of the IBE system. The most famous one is FIBE~\cite{sahai2005fuzzy}, which brings the concept of fuzziness to IBE. It regards identification features and verification features as two identity sets and defines a threshold value. When the intersection of the two identity sets exceeds the threshold, the authentication passes. 

Some other BBE schemes are as follows.
Guo et al.~\cite{guo2014poster} proposed a Euclidean distance-based method for biometric-based encryption. Their way of handling the inexact matching of biometric 
templates is different from ours. Guo et al.~\cite{guo2015distance} introduced Mahalanobis distance to encrypt the distance of biological templates. However, the calculation of Mahalanobis distance is complex and contains square root operations, which will reduce the accuracy of data that cannot be completely square rooted.
Carey et al.~\cite{carey2020cancelable} developed a biometric-based encryption scheme for transmitting medical images. Their scheme is based on AES, while our scheme is based on IBE.

\section{Proposed Scheme}

As mentioned in the introduction, the gait-based scheme AmSoBe consists of three components: ABLSTM, SOT, and BBE. 
The first two components are used for gait identification, while the last component is used for secure message interactions.

\subsection{Gait Identification Process}
\label{sec:gbip}
The gait identification process can be divided into the registration and identification stages, as shown in Fig.~\ref{sbt}.

During the registration stage, the owner uses the SWD to collect gait signals and extract gait features. Then the SWD encrypts the extracted gait feature into a reference gait template and transmits it to cloud storage.

During the identification stage, the owner uses SWD to collect gait signals and extract gait features. Afterward, the SWD encrypts the gait signal into an identification gait template and transmits it to the cloud storage. 

Finally, the owner is given the result of the cloud storage's comparison of the reference template and the identification gait template. To safeguard the owner's privacy, both templates are encrypted in the cloud storage.

\begin{figure}[htb]
\centering
\includegraphics[width=3.4in]{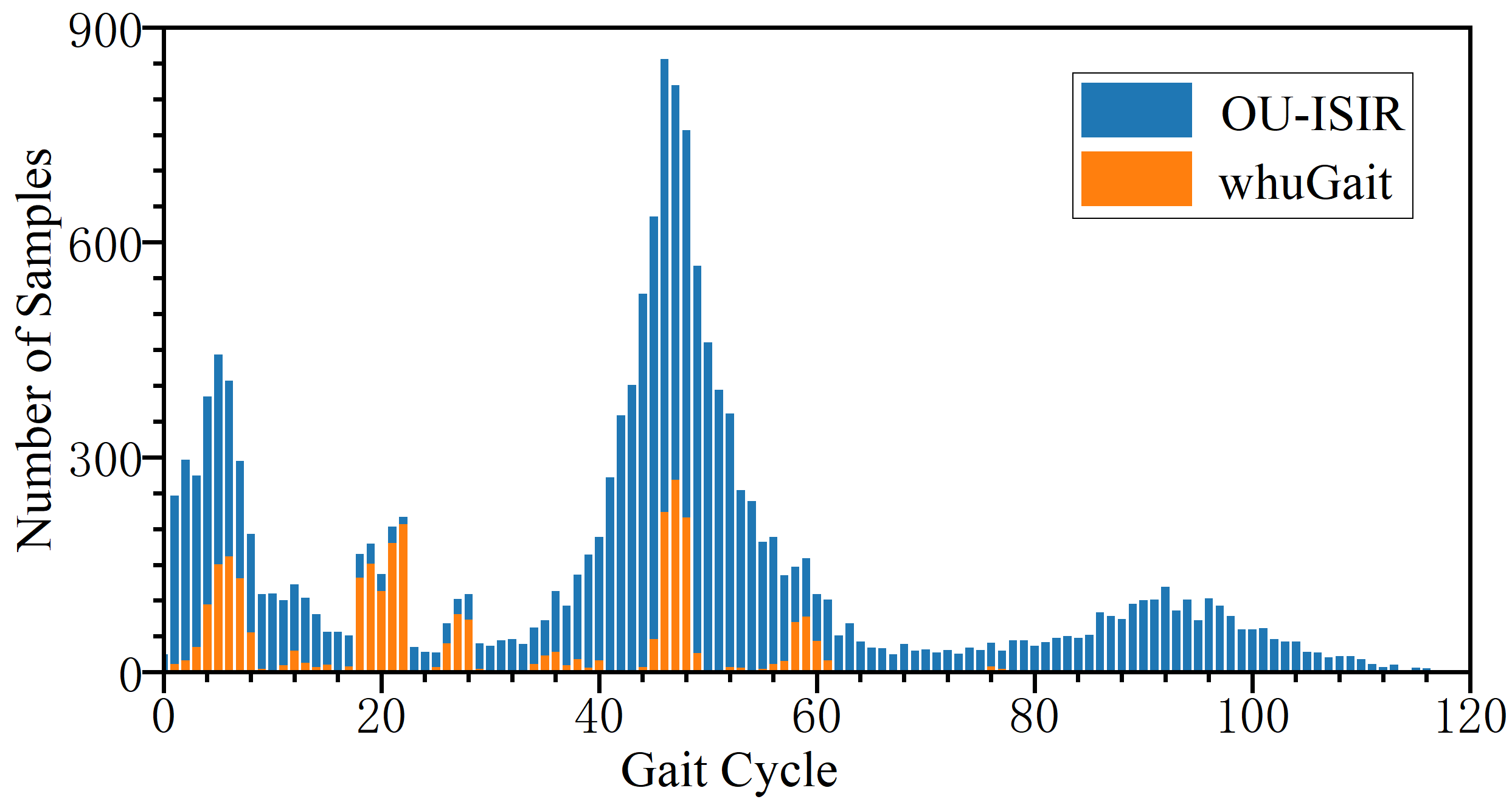}
\caption{Gait cycles distribution of the OU-ISIR and the whuGait databases.}
\label{cycle}
\end{figure}

\subsection{The ABLSTM Network}
\label{sec:ab}
In the identification process, the ABLSTM network is used as a feature extractor. This section introduces the dynamic cycle algorithm and the structure of the ABLSTM network.

%\subsection{Dynamic Cycle Algorithm}
%\label{sec:gsp}
The gait signals $S$ consist of the acceleration and gyroscope signals and contain 6 channels. The gait signals are represented as follows,
\begin{equation}
\label{eq:1}
S = (\mathbf{s}_{1},\mathbf{s}_{2},\mathbf{s}_{3},\mathbf{s}_{4},\mathbf{s}_{5},\mathbf{s}_{6}),
\end{equation}
where $\mathbf{s}_{1},\mathbf{s}_{2},\mathbf{s}_{3} \in  \mathbb{R}^{L}$ denote three-axis acceleration signals, $\mathbf{s}_{4},\mathbf{s}_{5},\mathbf{s}_{6} \in\mathbb{R}^{L}$ denote three-axis gyroscope signals, and $L$ is the length of the IMU sample vector.

The signals $S$ are processed in two steps. First, the wavelet transform is used to remove the noise from the signal. Second, $S$ is segmented according to the gait cycle calculated by our proposed dynamic cycle algorithm (DCA).

\begin{figure}[htb]
\centering
\includegraphics[width=3.6in]{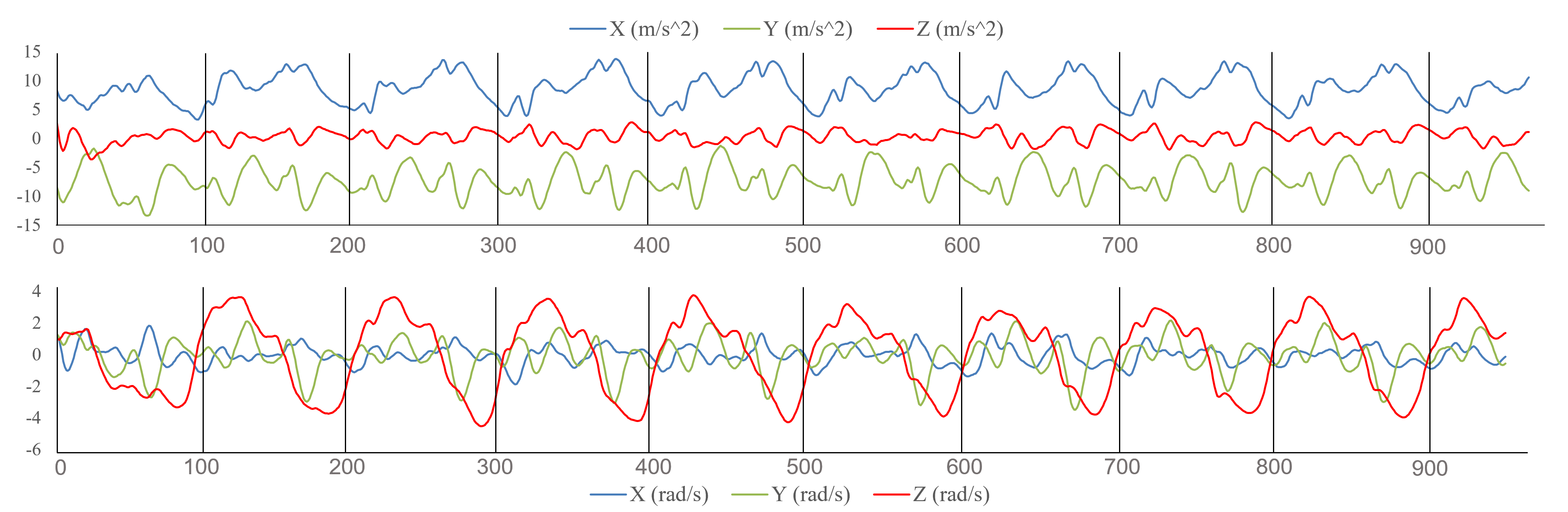}
\caption{An example segmentation of a 6-channel gait signal. The upper (lower) part corresponds to three-axis acceleration (gyroscope) signals.}
\label{gait}
\end{figure}

The gait cycle was previously calculated by the DTW algorithm or walking frequency. Those methods extend the signal's length by interpolation or reduce the gait length by down-sampling. They can introduce noise or reduce information about the original gait. The gait cycle calculated by walking frequency alone is not universal for large gait data. Our algorithm DCA solves those problems by using the Fourier transform to transform the gait signal into a superposition of sinusoidal signals with different amplitudes and cycles $T_{n}$. Considering the six-channel signals are all sampled by the same IMU, it suffices to calculate the cycle of channel $\mathbf{s}_{1}$. $\mathbf{s}_{1}$ is viewed as the output of the function $s(x)$ at every sampling time and fitted to the Fourier transform as a superposition of sinusoidal functions, expressed as follows,
\begin{equation}
\label{eq:3}
s(x) = {a}_{0} + \sum_{n=1}^\infty {a}_{n} sin(\frac{2\pi}{{T}_{n}}x), 
\end{equation}
where ${a}_{0}$ is the Fourier bias coefficient, and ${a}_{n}$ is the Fourier coefficient.
When there are enough sinusoidal signals, the original gait signal can be approximated as a superposition of sinusoidal signals. Each sinusoidal signal corresponds to a coefficient ${a}_{n}$. The larger the coefficient, the greater the cycle of the corresponding sinusoidal signal can be approximated as the gait cycle $T$, expressed as 
\begin{equation}
\begin{aligned}
    a &= \textup{max}({a}_{n}),\\
    &T \approx {T}_{a}.
\end{aligned}
\end{equation}

After splitting, each channel can be represented as
\begin{equation}
\begin{aligned}
    \mathbf{s}_{i} &= [\mathbf{x}_{i}^{1};~\mathbf{x}_{i}^{2};~\dots;~\mathbf{x}_{i}^{l}],  l = \lfloor {L}\big/{T} \rfloor,\\
     \mathrm {w.r.t.},~ \mathbf{x}_{i}^{j}&=(s_{i}^{(j-1)T+1};~s_{i}^{(j-1)T+2};~\dots;~s_{i}^{jT}),
\end{aligned}
\end{equation}
where $i \in [1, 6]$, $j \in [1, l]$, and $\lfloor \star \rfloor$ is the floor function.

To examine the length of a typical gait cycle, cycle statistics are conducted on the OU-ISIR and the whuGait databases, and the result is shown in Fig.~\ref{cycle}. Here, blue represents the statistics of the number of samples in each gait cycle of the OU-ISIR dataset and orange represents the whuGait dataset. The average gait cycle of the OU-ISIR dataset is 49.50 and the average gait cycle of the whuGait dataset is 57.58. According to the statistics, $T$ is set to $100$ to ensure that most input gait signal segments contain at least one complete gait cycle. An example of the segmentation of a six-channel gait signal is shown in Fig.~\ref{gait}.

 \textbf{Remark.} The gait cycle is an important feature that can help to recognize different people. In our gait recognition model, this information is ignored and the gait signals are segmented with a fixed interval $T=100$. It is interesting to explore whether adding the gait cycle as a feature to the gait recognition process can further increase the gait recognition accuracy, which is left as future work.

\begin{figure}[h]
\centering
\includegraphics[width=3.2in]{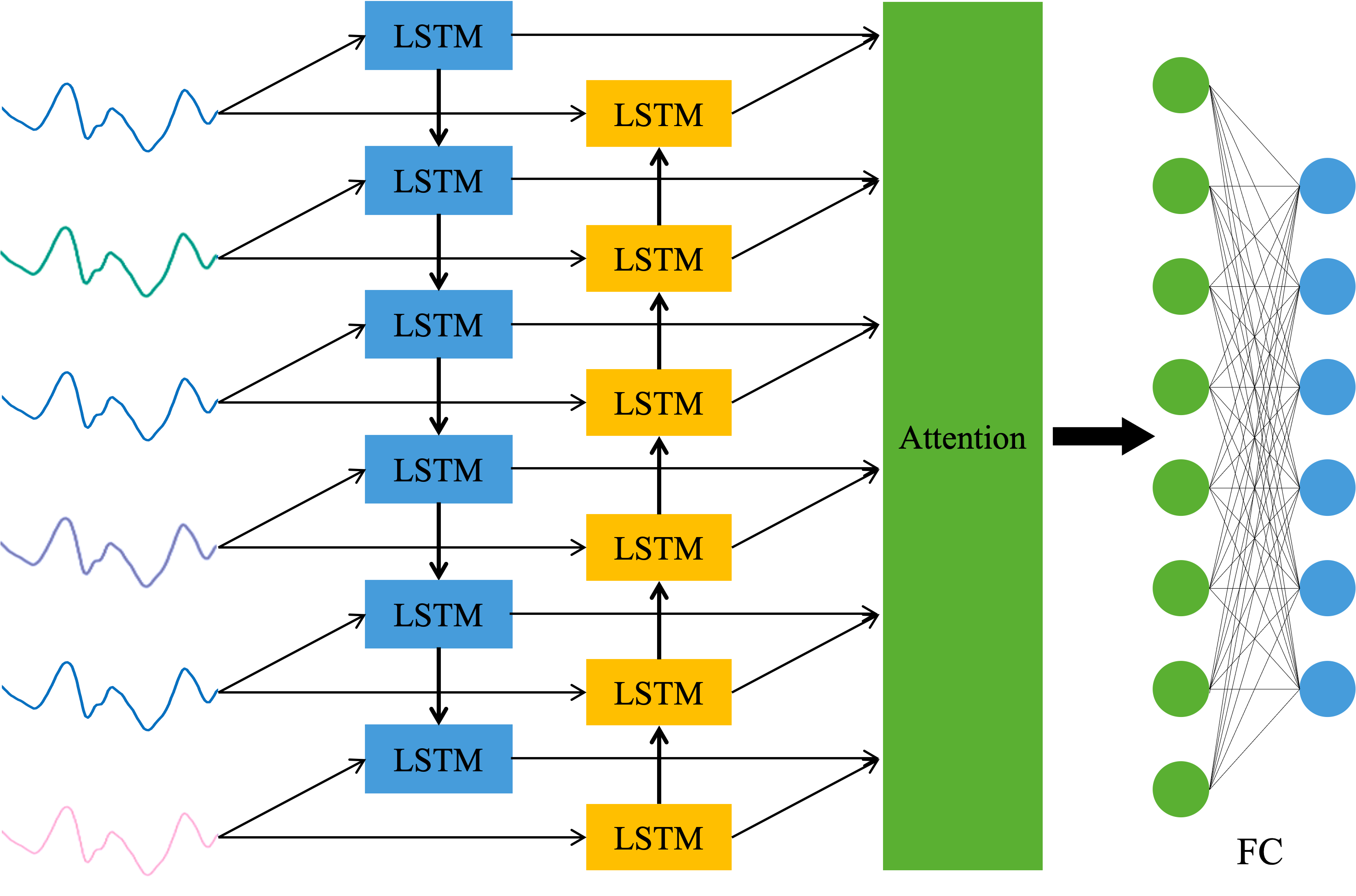}
\caption{The structure of the ABLSTM network.}
\label{ablstm}
\end{figure}

%\subsection{ABLSTM}
After the segmentation, the gait signals are input into the ABLSTM network for gait recognition. 
Figure~\ref{ablstm} shows the whole structure of the ABLSTM network. The ABLSTM network consists of the BLSTM layer and the attention layer. They can extract features $\mathbf{f}_{blstm}$ and $\mathbf{f}_{att}$ respectively. After the attention layer, a fully connected (FC) layer is added. It uses the feature vector $\mathbf{f}_{att}$ as its input. 

\subsubsection{BLSTM}
The BLSTM consists of a forward LSTM layer ($\overrightarrow{\textup{LSTM}}$) and a backward LSTM layer ($\overleftarrow{\textup{LSTM}}$). For one of the channels, the BLSTM input is the channel $\mathbf{s}_{i}$, the hidden state $\mathbf{h}_{0}$ and cell state $\mathbf{c}_{0}$.  $\mathbf{h}_{0}$ and $\mathbf{c}_{0}$ are randomly initialized at the beginning of network training.

An LSTM cell has three gates for protecting and controlling the state of each gate. At time $t$, the updates of each gate and state of the LSTM cell are as follows,
% the forget gate $f$, input gate $i$, and cell state $c$ can be updated as
\begin{equation}
\begin{aligned}
 \tilde{\mathbf{c}}_{t} &= \textup{tanh}({W}_{ic}   \mathbf{x}_{i}^{j} + {U}_{hc}   \mathbf{h}_{t-1} +  \mathbf{b}_{ic}), \\  
 \mathbf{c}_{t} &=  \mathbf{f}_{t} \circ  \mathbf{c}_{t-1} +  \mathbf{i}_{t} \circ   \tilde{\mathbf{c}}_{t},\\
  \mathbf{f}_{t} &= \sigma ( {W}_{if}  \mathbf{x}_{i}^{j} +  {U}_{hf}  \mathbf{h}_{t-1} +  \mathbf{b}_{if}), \\  
 \mathbf{i}_{t} &= \sigma ( {W}_{ii}  \mathbf{x}_{i}^{j} +  {U}_{hi}   \mathbf{h}_{t-1} +  \mathbf{b}_{ii}),\\
  \mathbf{o}_{t} &= \sigma ( {W}_{io}  \mathbf{x}_{i}^{j} +  {U}_{ho}   \mathbf{h}_{t-1} +  \mathbf{b}_{io}),\\
  \mathbf{h}_{t} &=  \mathbf{o}_{t} \circ \textup{tanh}( \mathbf{c}_{t}),
 \label{eq8}
 \end{aligned}
 \end{equation}
where $\mathbf{c}_{t} \in \mathbb{R}^{h}$ is the cell state, $\tilde{\mathbf{c}}_{t} \in \mathbb{R}^{h}$ is the activation vector at time $t$, $\mathbf{f}_{t}$, $\mathbf{i}_{t}$, $\mathbf{o}_{t} \in \mathbb{R}^{h}$ represent the forget, input and output gates at time $t$, $\mathbf{h}_{t}$ is the hidden state at time $t$, $\mathbf{h}_{t} \in \mathbb{R}^{h}$, $\sigma$ is a sigmoid function, $\circ$ is the Hamadard product, $W$ and $U$ are the weights of the corresponding gates, and $\mathbf{b}$ is the bias of the corresponding gates.

\begin{figure}[htb]
\centering
\includegraphics[width=3.2in]{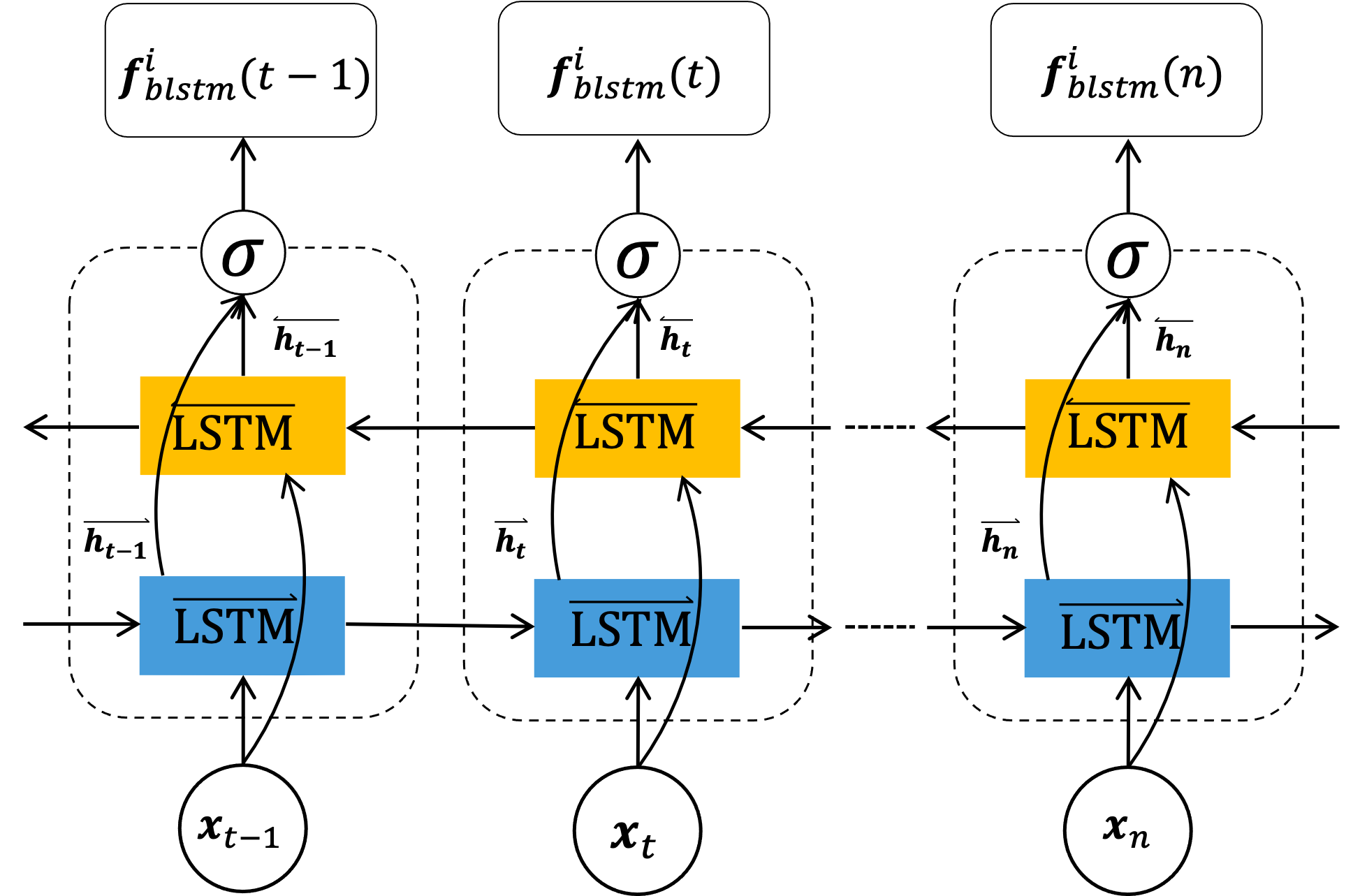}
\caption{The structure of the BLSTM.}
\label{blstm}
\end{figure}

The structure of the BLSTM is shown in Fig.~\ref{blstm}. At time $t$, the BLSTM concatenates the hidden layer states $\overrightarrow{\mathbf{h}_{t}}$ and $\overleftarrow{\mathbf{h}_{t}}$ of the $\overrightarrow{\textup{LSTM}}$ and $\overleftarrow{\textup{LSTM}}$ as the output $\mathbf{f}_{blstm}^{i} (t)$. The output $\mathbf{f}_{blstm} ^ {i}$ of BLSTM is expressed as
\begin{equation}
\begin{aligned} 
&\mathbf{f}_{blstm} ^ {i} = [\mathbf{f}_{blstm} ^ {i}(1);~\mathbf{f}_{blstm} ^ {i}(2);~...~;~\mathbf{f}_{blstm} ^ {i}(n)],\\
&\mathrm {w.r.t.},~{\mathbf{f}_{blstm} ^ {i}(t)} =\sigma [\overrightarrow{\mathbf{h}_{t}};~ \overleftarrow{\mathbf{h}_{t}}],
\label{eq14}
\end{aligned}
 \end{equation}
where $\mathbf{f}_{blstm} ^ {i} \in \mathbb{R}^{2nh}$, $\mathbf{f}_{blstm} ^ {i}(t) \in \mathbb{R}^{2h}$.

Finally, the output of the 6-channel BLSTM is concatenated into a vector as the feature vector $\mathbf{f}_{blstm}$, which is expressed as
\begin{equation}
    \mathbf{f}_{blstm} = [\mathbf{f}_{blstm} ^ {1}; \mathbf{f}_{blstm} ^ {2}; \ldots; \mathbf{f}_{blstm} ^ {6}],
\label{eq16}
\end{equation}
where $\mathbf{f}_{blstm} \in \mathbb{R}^{12nh}$.

\subsubsection{Attention}
Only left-to-right or right-to-left sequential computations are possible with LSTM networks. These networks have two issues. (1) The current output results depend on the results of the previous time. (2) There is information loss in the calculation process. Although the structure of gate mechanisms such as the BLSTM somewhat reduces the issue of long-term dependence, it remains ineffective for extremely long-term dependence. The attention mechanism, on the other hand, can effectively solve these two problems. The scaled dot-product attention is used~\cite{vaswani2017attention}, which is expressed as
\begin{equation} 
\mathbf{f}_{att} = \textup{softmax} ( {\frac{\mathbf{q} \mathbf{k}^{T}}{\sqrt{d_{k}}}}) \mathbf{v},
\label{eq17}
\end{equation} 
where $\mathbf{f}_{att} \in \mathbb{R}^{12nh}$, $\mathbf{q}$, $\mathbf{k}$, $\mathbf{v}$  represent query vector, key vector, and value vector respectively, and $\sqrt{d_{k}}$ is a scaling factor. When the order of magnitude is large, scaling with $\sqrt{d_{k}}$ prevents the softmax from assigning almost the entire probability distribution to the label corresponding to the maximum value. Additionally, it prevents the gradient from vanishing during backpropagation. The $\mathbf{q}$, $\mathbf{k}$, $\mathbf{v}$ and $d_{k}$ can be expressed as
\begin{equation}
\begin{aligned}
 \mathbf{q} &= W_{q} \mathbf{f}_{blstm},\\
 \mathbf{k} &= W_{k} \mathbf{f}_{blstm},\\  
 \mathbf{v} &= W_{v} \mathbf{f}_{blstm},\\
 d_{k} &=  \textup{Var}(\mathbf{q}^{T} \mathbf{k}),
\label{eq18}
\end{aligned}
\end{equation} 
where $W_{q}, W_{k}, W_{v} \in \mathbb{R}^{12nh \times 12nh}$ are weight matrices, $\mathbf{q}, \mathbf{k}, \mathbf{v} \in \mathbb{R}^{12nh}$, and $d_{k}$ is the variance of $\mathbf{q}^{T} \mathbf{k}$.

\subsubsection{Classifier}
After the attention layer, the FC layer is used as the classifier. The output $\mathbf{o}_{fc}$ of the FC layer is transformed by the sigmoid function to obtain $\Hat{\mathbf{y}} = \begin{bmatrix}  \Hat{y}_{1} &  \Hat{y}_{2} ~\ldots ~ \Hat{y}_{n_{c}}\end{bmatrix}$, where $n_{c}$ represents the number of all candidate subjects and $\Hat{y}_{i}$ ($1\leq i\leq n_{c}$) represents the probability of belonging to the $i$-th candidate subjects. After one-hot encoding, the labels for real subjects in the dataset are denoted as $\mathbf{y} = \begin{bmatrix} {y}_{1} & {y}_{2} ~\ldots ~{y}_{n_{c}}\end{bmatrix}$. The optimizer of the network uses the cross-entropy function as a loss function, which is expressed as follows,
\begin{equation}
\begin{aligned}
    \mathbf{f}_{att}^{'} &=  \textup{Dropout}(\mathbf{f}_{att}, p),\\
    \mathbf{o}_{fc} &= W_{fc}\mathbf{f}_{att}^{'} + \mathbf{b}_{fc},\\
    \Hat{\mathbf{y}} &= \sigma(\mathbf{o}_{fc}),\\
    \textup{CrossEntropyLoss} &= -\sum_{i=1}^{n_{c}} y_{i} \log (\frac{e^{\hat y_{i}}}{\sum_{j=1}^{n_{c}}e^{\hat y_{j}}}),
\label{eq23}
\end{aligned}
\end{equation} 
where $\mathbf{o}_{fc} \in \mathbb{R}^{n_{c}}$, $W_{fc} \in \mathbb{R}^{n_{c} \times 12nh}$, $\mathbf{b}_{fc} \in \mathbb{R}^{n_{c}}$, $p$ is a parameter of dropout that represents the probability that the elements of $\mathbf{f}_{att}$ are zeroed, $W_{fc}$ is the weight of the FC, and $\mathbf{b}_{fc}$ is the bias of the FC.

\subsection{The SOT Scheme}
\label{sec:s}
The gait feature privacy protection is essential to the recognition network because gait feature data is extremely private and cannot be undone once disclosed.

Two cases are envisaged.
In the first case, the SWD uses gait data for identification and the attacker obtains unencrypted gait data from cloud storage. The attacker can then directly input the gait data and pass the identification instead of the owner. In the second case, the SWD extracts gait features through the recognition network and performs identification and the attacker obtains unencrypted gait features and network parameters from cloud storage. In this case it is difficult for the attacker to pass the identification, because there are operations such as normalization and dropout (e.g., Equations~\eqref{eq8}-\eqref{eq23}) in the recognition network, which makes it difficult to reverse the computation from the output to obtain the input. When the attacker is assumed to get higher privileges to perform template matching directly, as in the first case, the attacker will be able to pass the identification.

The proposal of the SOT scheme enhances the security of identification. Even in the second case, an attacker bypassing the network for identification would be rejected.
As the input of the SOT scheme, the feature vector  ($\mathbf{f}_{att}$) is taken as the gait template. The reference templates $\mathbf{t}_{ref}$ and identification templates $\mathbf{t}_{idf}$ are expressed as follows,
\begin{equation}
\begin{aligned}
 \mathbf{t}_{ref} &= ({t}_{r1},{t}_{r2},\ldots,{t}_{rn}),\\
 \mathbf{t}_{idf} &=({t}_{s1},{t}_{s2},\ldots,{t}_{sn}).
\end{aligned}
\end{equation}

For later usage, the biological template distance $D$ between two vectors {\bf x} and {\bf y} is defined as
\begin{equation}
    D = \Vert\mathbf{x}-\mathbf{y}\Vert_{2}^{2} = \sum_{i=1}^n (x_i - y_i)^2,
\end{equation}
where $\mathbf{x},~\mathbf{y} \in \mathbb{R}^n, ~D \in [0, 1]$, and $\Vert \cdot \Vert_2$ is the 2-norm.
Let $\bf x'$ and $\bf y'$ be two $(n + 2)$-length vectors, which are expressed as follows, 
\begin{eqnarray}
 {\bf x'}&=&(x_{1}, x_{2}, \cdots , x_{n},1, \sum \limits _{i = 1}^{n} x_i^2),\label{eq:29}\\
 {\bf y'}&=& (-2y_{1}, -2y_{2}, \cdots , -2y_{n}, \sum \limits _{i = 1}^{n} y_i^2,1).\label{eq:30}
\end{eqnarray}
The biological template distance $D$ can then be written as
\begin{equation}
    D = \langle {\bf x'}, {\bf y'} \rangle,
\end{equation}
where $\langle \cdot , \cdot \rangle$ is the inner product.

The SOT scheme is divided into three steps: biometric preprocessing, biometric template encoding, and biometric template decoding~\cite{li2012ear}. Biometric template encryption is commonly performed directly on the biometric template, such as binding the key to the biometric template~\cite{uludag2004biometric}, fuzzy vault~\cite{5973854}, and matrix transformation-based encryption method~\cite{liu2019efficient}. Considering the security of encryption and the reliability of experiments, a stochastic orthogonal transformation encryption scheme is proposed, including the following steps.

1) The process of encrypting the reference template $\mathbf{t}_{ref}$. The SWD expands the template according to Eq.~\eqref{eq:29} to $\mathbf{t}_{ref}^{''}$, 
randomly selects a position $p \in \{1,2, \cdots, n+3 \}$, and inserts a positive constant $a$ before the $p$-th entry of $\mathbf{t}_{ref}^{''}$. For $p=n+3$, the resulting vector
can be expressed as follows,
\begin{equation}
\mathbf{t}_{ref}^{'} = ({t}_{r1},{t}_{r2},\ldots,{t}_{rn}, 1, \sum \limits _{j= 1}^{n} t_{rj}^2, a ).
\label{eq:31}
\end{equation}
As another example, for $p=1$, the resulting vector is $ \mathbf{t}_{ref}^{'} = (a, {t}_{r1},{t}_{r2},\ldots,{t}_{rn}, 1, \sum_{j} t_{rj}^2  ).$

Then, the SWD randomly generates an orthogonal matrix $M$ to encrypt the extended template $\mathbf{t}_{ref}^{'}$ and transmits the encrypted reference template $\mathbf{t}_{enc}$ to the cloud storage, which can be expressed as
\begin{equation}
\mathbf{t}_{enc} = \alpha\mathbf{t}_{ref}^{'} \times M,
\label{eq:32}
\end{equation}
\noindent where $M \in \mathbb{R}^{{(n+3) \times (n+3)}}$, $\alpha$ is a random positive real number, and $\times$ is the matrix multiplication.

2) The process of encrypting the identification template $\mathbf{t}_{idf}$. The SWD expands the template according to Eq.~\eqref{eq:30} to $\mathbf{t}_{idf}^{''}$, and inserts a negative constant $b$ before the $p$-th entry of $\mathbf{t}_{idf}^{''}$ using the secret key $p$. For $p=n+3$, the resulting vector can be expressed as follows,
\begin{equation}
\mathbf{t}_{idf}^{'} = (-2{t}_{s1},-2{t}_{s2},\ldots,-2{t}_{sn},\sum \limits _{j = 1}^{n} t_{sj}^2,1, b).
\label{eq:33}
\end{equation}

Subsequently, the SWD uses the secret key $M$ to encrypt the extended template $\mathbf{t}_{idf}^{'}$ and transmits the encrypted identification template $\mathbf{t}_{enc}^{'}$ to the cloud storage, which can be expressed as
\begin{equation}
\mathbf{t}_{enc}^{'} = \alpha\mathbf{t}_{idf}^{'} \times M.
\label{eq:34}
\end{equation}

3) Identification stage. Our scheme runs a matching algorithm in cloud storage, computing the biological template distance between two encrypted templates, which is expressed as follows,
\begin{equation}
    \langle \mathbf{t}_{enc}, \mathbf{t}_{enc}^{'} \rangle = \alpha^{2}(D+ab),
\label{eq:35}
\end{equation}
where $ab$ represents the matching threshold. When $\langle \mathbf{t}_{enc}, \mathbf{t}_{enc}^{'} \rangle \le 0,$ the two templates are similar and the identification is successful. Conversely, the templates are not similar and the identification fails. Because biometrics are subject to noise interference when sampling, thresholds can better solve the fuzzy problem of biometrics.

{\bf Remark.} The SOT encryption scheme presented here can be slightly extended. Instead of inserting just one parameter $a$ into the reference template, multiple positive parameters $a_1, a_2, \cdots, a_m$ can be inserted at random positions. Multiple negative parameters $b_1, b_2, \cdots, b_m$ are then 
inserted into the identification template at corresponding positions. In later evaluations, the effect of inserting different numbers of parameters will be examined.

\subsection{The BBE Scheme}
\begin{figure}[htb]
\centering
\includegraphics[width=8.5cm]{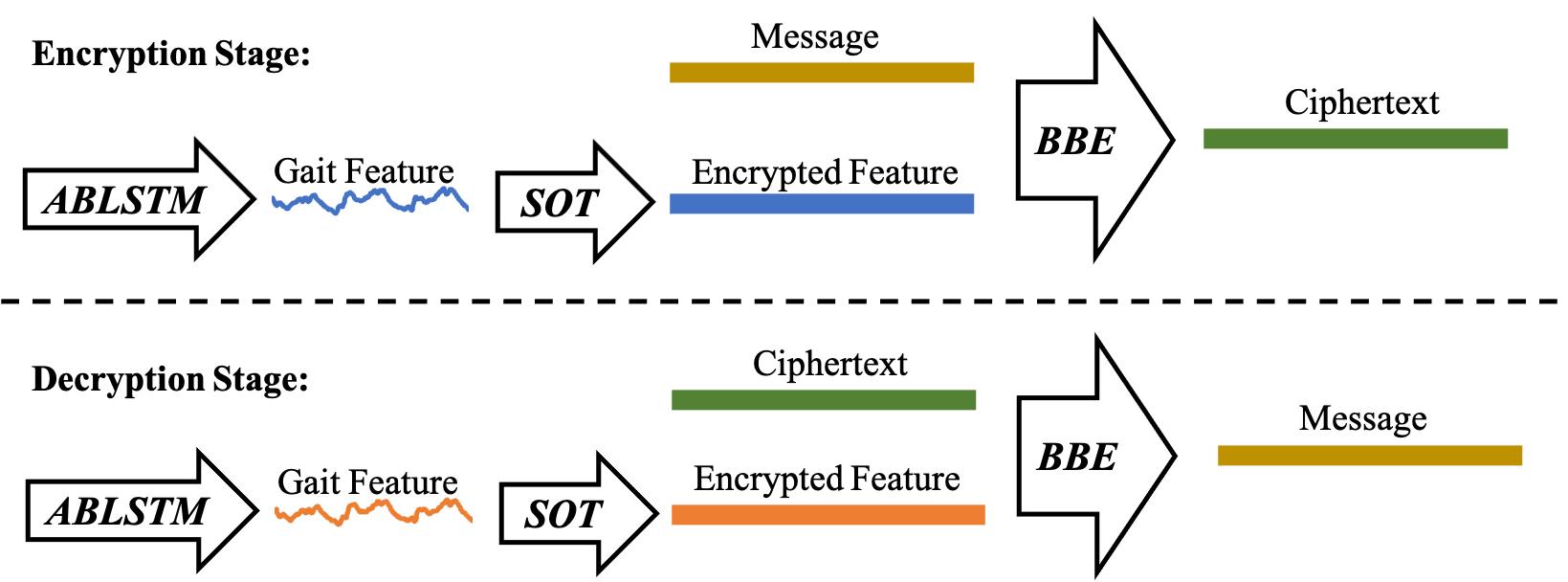}
\caption{The overall architecture of the BBE scheme.}
\label{fig:BBEillus}
\end{figure}

A BBE scheme is proposed to enable secure message interactions between SWD and other devices. The BBE scheme is based on the idea of IBE, and is integrated with the SOT encryption method to make it have biometric fault tolerance and good encryption performance.
The relations between ABLSTM, SOT, and BBE are illustrated in Fig.~\ref{fig:BBEillus}.

%\subsection{Definitions}
Let us start by defining BBE. Different from the IBE model, the BBE scheme uses biological template distance to calculate the similarity of two biological templates. For security, the BBE scheme is similar to the IBE's selective ID scheme~\cite{boneh2004efficient}. The game between the attacker and the challenger in the implementation of the BBE encryption scheme is introduced as follows.

\begin{enumerate}[\labelwidth=35pt]
\item [\textbf{Initializing}] The biometric $\mathbf{w}$ has been declared to be under challenge by the adversary $\mathcal{A}$.
\item[\textbf{Stage 1}] The challenger generates the key and provides the public parameters to $\mathcal{A}$.
\item[\textbf{Stage 2}] $\mathcal{A}$ has permission to check private keys of several biometric, $\mathbf{z}_i$, where $\langle \mathbf{w}, \mathbf{z}_i \rangle = 0$ for all $i$. The challenger generates the private key. Then the adversary receives the private key from the challenger.
\item[\textbf{Challenging}] The challenger receives the same length message, $I_{0}$ and $I_{1}$, from the adversary. The challenger then randomly selects an element $\mu \in \{0, 1\}$, computes the corresponding message $I_{\mu}$, and sends it to $\mathcal{A}$.
\item[\textbf{Stage 3}] Repeat Stage 2.
\item[\textbf{Guessing}] If $\mathcal{A}$ guesses the original of the challenger encryption messages correctly, $\mu' = \mu$, $\mathcal{A}$ wins the game. If not, $\mathcal{A}$ loses the game.
\end{enumerate}

\textbf{Definition~2.} When the advantage for all attackers $\mathcal{A}$ in polynomial time is negligible, the BBE method is defined as secure, which is expressed as follows,
\begin{equation}
\text {Adv}_{\mathcal {A}(\lambda)}^{BBE}= \left | \Pr [{\mathrm{ \mu}}'=\,{\mathrm{ \mu}}]-\frac {1}{2} \right | \le \textsf{negl}(\lambda),
\end{equation}
where $\lambda$ represents the bit number of the group size $p$.

In addition, the following related knowledge is needed in the construction of BBE.

\subsubsection{Bilinear Maps}
Let $\mathbb {G}$ and $\mathbb {G}_{T}$ be the elliptic subgroup and multiplicative subgroup of prime order $p$, respectively. Then $\mathbb{G}\times \mathbb{G}\rightarrow \mathbb{G}_{T}$ is a bilinear map if the following two conditions hold:

1. ${e}(g^{\varphi_{1}}, g^{\varphi_{2}})={e}(g, g)^{\varphi_{1}\varphi_{2}}$.

2. ${e}(g, g)\neq 1_{\mathbb{F}}$, where $g$ is a generator of $\mathbb {G}$.

\subsubsection{DBDH assumption}
Given a bilinear mapping and random numbers $\varphi_{1},\varphi_{2},\varphi_{3}\in \mathbb{Z}_{p}$. $\mathcal{A}$ receives the parameters $\Theta_0=\{g, g^{\varphi_{1}}, g^{\varphi_{2}}, g^{\varphi_{3}}, e(g, g)^{\varphi_{1}\varphi_{2}\varphi_{3}}\}$ and $\Theta_1 = \{g, g^{\varphi_{1}}, g^{\varphi_{2}}, g^{\varphi_{3}}, \phi\}$ and determines whether they are equal. Its advantage is expressed as 
\begin{equation}
    \text {Adv}_{\mathcal {A}(\lambda)}^{DBDH} = \bigg| \Pr[\mathcal{A}(\Theta_0)=1] -\Pr[\mathcal{A}(\Theta_1)=1] \bigg|. 
\end{equation}
The DBDH assumption is satisfied if for all polynomial time distinguisher $\mathcal{A}$, 
\begin{equation}
    \text {Adv}_{\mathcal {A}(\lambda)}^{DBDH} \le \textsf{negl}(\lambda),
 \end{equation}
where $\lambda$ represents the bit number of the group size $p$.

%\subsection{Construction}
%\label{bbecst}
With the related knowledge at hand, we are now ready to construct the BBE scheme by designing the following four main algorithms:

% Biometric-based encryption uses a reference biometric template {\bf w} to encrypt messages. The ciphertext can be decrypted correctly by the identification biometric template {\bf z} if and only if $D \leq d$, where $D$ is the biological template distance of {\bf w} and {\bf z}, $d$ is the threshold value. Biometric-based encryption comprises four algorithms.

{\bf Setup} The input is ${\bf w}=(w_{1},w_{2},\cdots ,~w_{n}) \in \mathbb {Z} ^{n}_{p}$. Then, the algorithm chooses random $\alpha _{i}, \beta$ from $\mathbb {Z}_{p}$ for all $i \in [1, n]$. The published public parameters are
$T_1 = g^{w_1}, \cdots, T_n=g^{w_n}, Y = e(g, g)^{\beta}, g_{i}=g^{\alpha_{i}}$. The master secret key is $(\alpha _{i}, \beta)$. 

{\bf Key Generation} The algorithm takes a $n$-length identification biometric template ${\bf z}=(z_{1},z_{2},\cdots ,z_{n})\in \mathbb {Z} ^{n}_{p}$ and the master key $(\alpha _{i}, \beta)$ as input. The algorithm randomly chooses $t \in \mathbb {Z}_{p}$. The private key consists of components, $(D, g^{t}) \in \mathbb {G} \times \mathbb {G}$, where $D=g^{\beta +t\sum _{i=1}^{n} \alpha _{i} z_{i}}$.

{\bf  Encryption} The algorithm randomly selects $r,s$ from $\mathbb {Z}_p$. The message $I$ is encrypted as follows,
\begin{equation}
    E = (E' = Y^{r}I, g^r, \{E_{i}=g_{i}^{r} T_{i}^{s}\}),
\end{equation}
\noindent which can also be written as 
\begin{equation}
    E = (C_I, C_0, \{C_1, \cdots, C_n\}).
\end{equation}

{\bf  Decryption} The $C_I$ is encrypted by the reference biological template {\bf w}. Only when the identification biological template {\bf z} satisfies ${\langle {\bf w}, {\bf z} \rangle}=0$, the $C_I$ can be decrypted by the private key. The decryption algorithm starts by calculating
\begin{equation}
\begin{aligned}
    e_{0}=&e\left({ g^{t}, \prod _{i=1}^{n}C_{i}^{z_{i}} }\right),\\ e_{1}=&e\left({ D, C_{0}}\right).
\end{aligned}
\end{equation}

When ${\langle {\bf w}, {\bf z} \rangle}=0$, the ciphertext can be decrypted as
\begin{equation}
    C_{I}\cdot e_{1}^{-1} e_{0} =I.
\end{equation}

%\subsection{Pre-processing with SOT}

In this BBE scheme, there is a necessary condition for the decryption algorithm, that is ${\langle {\bf w}, {\bf z} \rangle}=0$. To fulfill this condition, SOT is used to pre-process the biological templates, see Algorithm~\ref{alg1}. In this preprocessing algorithm, the SOT algorithm (Eqs.~\eqref{eq:31}-\eqref{eq:35}) is first run to compare the similarity of the two templates. If the identification template passes the similarity test, ${\bf z}$ is updated so that ${\langle {\bf w}, {\bf z} \rangle}=0$ holds. The details of the update can be found in Algorithm~\ref{alg1}. Then the SOT algorithm is run to encrypt the two templates into   ${\bf w'}$  and ${\bf z'}$. Finally the BBE scheme is run on the encrypted templates generated by SOT. This complete scheme is also called the SOT-BBE scheme.

\begin{algorithm}[tbh]
\caption{Preprocessing ${\bf w}$ and ${\bf z}$}
\label{alg1}
\begin{algorithmic}[1]
\REQUIRE~~\\
The biometric templates ${\bf w} = (w_{1}, w_{2}, \cdots , w_{n})$\\
The biometric templates ${\bf z} = (z_{1}, z_{2}, \cdots , z_{n})$ 
% \ENSURE The biometric templates {\bf w} and {\bf z}
\STATE
\STATE {\bf Initialize:}
\STATE ${\bf w} \Leftarrow (w_{1}, w_{2}, \cdots , w_{n}, 1, \sum \limits _{j = 1}^{n} w_j^2, a)$
\STATE ${\bf z} \Leftarrow (-2z_{1}, -2z_{2}, \cdots , -2z_{n}, \sum \limits _{j = 1}^{n} z_j^2, 1, b)$ 
\STATE $d \Leftarrow D+ab \Leftarrow SOT({\bf w}, {\bf z})$ 
\STATE
\STATE {\bf Update:}
\IF{$d \le 0$} 
\STATE ${\bf z} \Leftarrow (-2z_{1}, -2z_{2}, \cdots , -2z_{n}, \sum \limits _{j = 1}^{n} z_j^2, 1,  -D/a)$ 
\ELSE 
\STATE Pass
\ENDIF 
\STATE
\ENSURE The biometric templates ${\bf w}$ and ${\bf z}$
\end{algorithmic}
\end{algorithm}

Regarding the advantages of the SOT-BBE scheme over other existing BBE schemes \cite{guo2014poster,guo2015distance,carey2020cancelable}, the main advantage is on the protection of gait features. While other BBE schemes \cite{guo2014poster,guo2015distance,carey2020cancelable} ignore the protection of the biometric features when using it for encryption, the SOT-BBE scheme can protect the biometric feature from leakage during encryption.

\section{Security analysis}

The security analysis consists of two parts: security of SOT and reliability of BBE.

\subsection{Security of SOT}
\label{indcpa}
To define the CPA security of SOT, let us first describe the CPA experiment of SOT. Let
$\mathcal{A}$ represent the adversary and $\mathcal{E}$ the encryption scheme $(Dec,Enc)$. The challenger is represented by $\mathcal{C}$, i.e. the SWD. 
\begin{itemize}
% \itemindent 2.8em
\item[(1)] The $Gen(1^{\lambda})$ algorithm generates the secret key $sk$ and $\lambda$ is a security parameter.
\item[(2)] Given a \(1^{\lambda}\) input and oracle access to $Enc_{sk}(\cdot)$, $\mathcal{A}$ generates a pair of identically sized gait template plaintexts, $t_{0}$ and $t_{1}$.
\item[(3)] $\mathcal{C}$ selects bit $B\in\{0,1\}$ at random, and runs $Enc_{sk}(t_B)$ to get the ciphertext. Then, $\mathcal{C}$ sends the ciphertext to $\mathcal{A}$.
\item[(4)] $\mathcal{A}$ still gets $Enc_{sk}(\cdot)$ and outputs another bit $B^{'}$.
\item[(5)] If $\mathcal{A}$ guesses the original of the challenger encryption template correctly, namely $B=B^{'}$, $\mathcal{A}$ wins the game. If not, $\mathcal{A}$ loses the game.
\end{itemize}

The CPA security for the SOT scheme based on the CPA experiment is defined as follows.

\textbf{Definition~1.} When the advantage of all attackers $\mathcal{A}$ in polynomial time is negligible, the SOT scheme is defined to be CPA-secure, expressed as follows,
\begin{equation} 
\left | \Pr (\text {CPA}_{\mathcal {A}} (\lambda) = 1) - \frac {1}{2} \right | \le \textsf{negl}(\lambda).
\end{equation}

According to the CPA experiment, it is only necessary to prove whether the attacker $\mathcal{A}$ can know which original template is the encrypted gait template sent by the challenger $\mathcal{C}$ (SWD). In multiple encryptions, the scheme with CPA security is still secure~\cite{katz2020introduction}. So it suffices to prove security for the case that  the attacker $\mathcal{A}$ sends two templates $\mathbf{t}_0$ and $\mathbf{t}_1$ to the challenger $\mathcal{C}$.

The challenger $\mathcal{C}$ receives two same-length templates, 
$\mathbf{t}_0$ and $\mathbf{t}_1$, from the attacker $\mathcal{A}$. By simply comparing the lengths of encrypted templates, the attacker $\mathcal{A}$ cannot deduce which template was encrypted. Thus $\mathcal{A}$ still gets $Enc_{sk}(\cdot)$ and knows the two templates $\mathbf{t}_0$ and $\mathbf{t}_1$, $\mathcal{A}$ can figure out the encrypted template $\mathbf{t}_{enc}^{0}\leftarrow Enc(sk,\mathbf{t}'_{0})$  and $\mathbf{t}_{enc}^{1}\leftarrow Enc(sk,\mathbf{t}'_{1})$ by itself. In the CPA experiment, the attacker $\mathcal{A}$ computes the encrypted templates $\mathbf{t}_{enc}^{0}$ and $\mathbf{t}_{enc}^{1}$ by having oracle access to $Enc(sk,\cdot)$. Then $\mathcal{A}$ compares them with the encrypted templates sent by the challenger $\mathcal{C}$. $\mathcal{A}$ guesses which template the challenger has encrypted. However, this approach does not work because the parameters in the encryption scheme are random.

The encrypted template $\mathbf{t}_{enc}^{i}$ is taken into account. $\mathbf{t}_i$ is extended according to Eq.~\eqref{eq:31}, which is expressed as follows,
\begin{equation}
\mathbf{t}_{i}^{'} = ({t}_{i1},{t}_{i2},\ldots,{t}_{in}, 1, \sum \limits _{j = 1}^{n} t_{ij}^2, a).
\end{equation}
Then the encrypted templates $\mathbf{t}_{enc}^{i} = \alpha_{i}\mathbf{t}'_{i}M$ is obtained, where the element can be represented as
\begin{equation}
\begin{aligned}
{{t}_{enc}^{ij}}=&{\alpha _{i}}\sum \limits _{k = 1}^{n} {{t_{ik}}{m_{kj}}}+{\alpha _{i}} m_{(n+1)j}+{\alpha _{i}} m_{(n+2)j}\sum \limits _{j = 1}^{n} t_{ij}^2\\&+ {\alpha _{i}}m_{(n + 3)j}a. 
\end{aligned}
\label{eq:38}
\end{equation}

In Eq.~\eqref{eq:38}, $\alpha_{i}$ ($i=0,1$) are randomly selected parameters and are updated again for each encryption. The position of $a$ is also randomly selected and related to the template length $n$ and the probability when the two templates are equal is $\frac{1}{n+3} \times \frac{1}{2}\times(2^{-\lambda})^{2}=\frac{1}{n+3}2^{-2\lambda-1}$, in which $n+3$ is the probability of inserting $a$ at different positions of the template. In the experiment, $\lambda$ is set to $128$ and $n$ is set to $600$. So the advantage is about $2^{-266}$, which can be expressed as follows,
\begin{equation}
\left |{ \Pr (\text {CPA}_{\mathcal {A}} (\lambda) = 1) - \frac {1}{2} }\right | \le \textsf{negl}(\lambda).
\end{equation}

With the SOT encryption method, the user's gait template data is effectively protected and prevented from leakage that could lead to permanent template disclosure. This also enhances the security of our identification network.

\subsection{Reliability of BBE}
This section analyzes whether the BBE scheme is reliable from two aspects: correctness and security.
\subsubsection{Correctness Analysis}
If ${\langle {\bf w}, {\bf z} \rangle}=0$ is satisfied, the ciphertext is decrypted as follow,
\begin{equation}
    \begin{aligned}
        C_{I}\cdot e_{1}^{-1} e_{0}=&E' \cdot e(g,g)^{-\beta r}\cdot e(g,g)^{-rt \sum _{i=1}^{n} \alpha _{i} z_{i}} \\[1.5pt]&\cdot \, e(g,g)^{rt\sum _{i=1}^{n} \alpha _{i}z_{i}+st\langle {\bf w}, {\bf z}\rangle }\\[1.5pt]=&e(g,g)^{\beta r} I \cdot e(g,g)^{-\beta r}\cdot e(g,g)^{st\langle {\bf w}, {\bf z}\rangle  }\\[1.5pt]=&I.
    \end{aligned}
\end{equation}

\subsubsection{Security Proof}
Assume that our construction is challenged by $\mathcal {A}$ who has the advantage $\text {Adv}_{\mathcal {A}(\lambda)}^{BBE}$ in the selective game. The DBDH issue is solved using the reduction algorithm $\mathcal {B}$, which has the advantage $\text {Adv}_{\mathcal {B}(\lambda)}^{DBDH}$. Given as input $(g, g^{\varphi_{1}}, g^{\varphi_{2}}, g^{\varphi_{3}}, \phi)$ from a pairing group $\mathbb {G}_T$ and play the security game as demonstrated below.

% \begin{enumerate}
\noindent \textbf{Initializing} The $n$-length vector ${\bf {w}^{*}}=(w_{1}^{*}, w_{2}^{*}, \cdots , w_{n}^{*})\in \mathbb {Z} _{p}^{n}$ has been declared to be under challenge by $\mathcal{A}$.

\noindent\textbf{Stage 1} The challenger generates the master secret key $(\alpha _{i}, \beta)$ as follows,
\begin{equation} 
\alpha _{i}=-\eta w_{i}^{*}\varphi_{2}+\eta _{i},\quad \beta =\varphi_{2}\varphi_{3}, 
\end{equation}
where $\eta ,\eta _{i}\in \mathbb {Z} _{p}$ are randomly  selected by the reduction algorithm $\mathcal {B}$; $\varphi_{2},\varphi_{3}$ are the parameters obtained after instantiating the DBDH problem.

The public parameters $(T_i = g^{w_i}, Y = e(g, g)^{\beta}, g_{i}=g^{\alpha_{i}})$ are simulated as follows,
\begin{equation}
\begin{aligned} 
g_{i}=&g^{\alpha _{i}}=g^{-\eta w_{i}^{*}\varphi_{2}+\eta _{i}}=(g^{\varphi_{2}})^{-\eta w_{i}^{*}}\cdot g^{\eta _{i}},\\[6pt] u=&e(g,g)^{\beta }=e(g^{\varphi_{2}}, g^{\varphi_{3}}). 
\end{aligned}
\end{equation}

The reduction algorithm uses random values $\eta,\eta _i$ to randomize the master secret key $\alpha _i,\beta$ and embeds the parameters $\varphi_{2},\varphi_{3}$ into the master public key. As a result, these parameters are independent of the adversary's perspective and uniformly random.

\noindent\textbf{Stage 2} $\mathcal{A}$ checks private keys of $n$-length biometric, ${\bf z}=(z_{1}, z_{2},~z_{3},\cdots ,z_{n})$, where $\langle {\bf {w}^{*}}, {\bf z} \rangle = 0$. $\mathcal {B}$ generates the private key $(sk_{1}, sk_{2})$, which is expressed as follows,
\begin{equation}
\begin{aligned} 
sk_{1}=&(g^{\varphi_{2}})^{-\eta t'\sum _{i=1}^{n}w_{i}^{*}z_{i}}(g^{\varphi_{3}})^{\frac {\sum _{i=1}^{n}\eta _{i}z_{i}}{ \eta }}g^{t'\sum _{i=1}^{n}\eta _{i}z_{i}},\\[5pt] sk_{2}=&(g^{\varphi_{3}})^{\frac {1}{ \eta }}g^{t'}, 
\end{aligned}
\end{equation}
where $t' \in \mathbb {Z}_{p}$ is selected by $\mathcal {B}$. The private key $(sk_{1}, sk_{2})$ is rewritten as
\begin{equation}
sk_{1}=g^{\beta +t\sum _{i=1}^{n}\alpha _{i} z_{i}}, \quad sk_{2}=g^{t}, 
\end{equation}
where $t=\frac {1}{ \eta }\varphi_{3}+t'$. Then the adversary receives the private key $(sk_{1}, sk_{2})$ from $\mathcal {B}$.

\noindent\textbf{Challenging} The challenger receives the same length messages, $I_{0},I_{1}$, from $\mathcal{A}$ to challenge on the vector ${\bf {w}^{*}}=(w_{1}^{*}, w_{2}^{*}, \cdots , w_{n}^{*})$. $\mathcal {B}$ creates the challenge ciphertext $E^{*}$, which is expressed as follows,
% The adversary returns $I_{0},I_{1}\in \mathbb {G} _{T}$ for challenge on the challenge vector ${\bf {w}^{*}}=(w_{1}^{*}, w_{2}^{*}, \cdots , w_{n}^{*})$. The reduction algorithm randomly chooses $i\in \{0,1\}$ and creates the challenge ciphertext $E^{*}$ as
\begin{equation}
\begin{aligned} 
E^{*}=&\big (C_{I},C_{0},C_{i}\big )\\=&\Big (\phi\cdot I_{\mu}, g^{\varphi_{1}},g^{\eta _{i}\varphi_{1}}\Big ), 
\end{aligned}
\end{equation}
where $\mu \in \{0, 1\}$ is randomly selected by $\mathcal {B}$ and $i \in (1, n)$.
If $\phi=e(g,g)^{\varphi_{1}\varphi_{2}\varphi_{3}}$, then
\begin{equation}
\begin{aligned} 
\phi\cdot I_{\mu}=&e(g,g)^{\beta r}\cdot I_{\mu},\\ 
g^{\eta _{i} \varphi_{1}}=&g^{(-\eta w_{i}^{*}\varphi_{2}+\eta _{i})\varphi_{1}+w_{i}^{*}\cdot \eta \varphi_{1}\varphi_{2}}\\
=&g^{\alpha _{i} r+w_{i}^{*} s}\\=&g_{i}^{r} g^{sw_{i}^{*}}, 
\end{aligned}
\end{equation}
where $r=\varphi_{1}$ and $s=\eta \cdot \varphi_{1} \cdot \varphi_{2}$. Finally the ciphertext is sent to $\mathcal {A}$.

\noindent\textbf{Stage 3} Repeat Stage 2.

\noindent\textbf{Guessing} $\mathcal {A}$ guesses $\mu' = \mu$. If the guess is correct, $\mathcal {B}$ returns 1 to guess $\phi=e(g,g)^{\varphi_{1}\varphi_{2}\varphi_{3}}$. Otherwise, $\mathcal {B}$ returns 0 to guess $\phi$ as a random group element.
% \end{enumerate}

The reduction proof is now complete. If $\phi=e(g,g)^{\varphi_{1}\varphi_{2}\varphi_{3}}$, a BBE ciphertext $E^{*}$ and $\mathcal {A}$ can guess $\mu$ with advantage $\text {Adv}_{\mathcal {A}(\lambda)}^{BBE}$. Otherwise, it can be inferred $\phi$ is an element in a random group. $\mathcal {A}$ considers $I_{\mu}$ is encrypted with disposable padding, so $\mathcal {A}$ guesses $\mu$ with no advantage. Therefore, $\text {Adv}_{\mathcal {A}(\lambda)}^{BBE}$ is the advantage for $\mathcal {B}$ to solve the hard DBDH problem. Each generation of the private key requires constant computation and no miscarriage occurs in our scenario.

\section{Experiment}

The experiments consists of three parts: the ABLSTM experiment, the SOT experiment, and the BBE experiment.

\subsection{ABLSTM Experiment}
Our ABLSTM network is built on the PyTorch 1.9.0 framework running on Python 3.8.8. The computer for our experiments consists of an Intel CPU (2.4GHz, 20 cores), a GeForce RTX3090 GPU (16G), and 128G of RAM. Table~\ref{table1} displays the parameter values, where $n_{c}\_ou$ and $n_{c}\_whu$ represent the corresponding $n_{c}$ values of the two datasets, respectively.

\begin{table}[htb]
\centering
\caption{The value of each parameter in the experiment.}
\begin{tabular}{ccccccc}
\hline
\bf L   & \bf T   & \bf n & \bf h  & \bf p   & \bf $\bf n_{c}$\_ou & \bf $\bf n_{c}$\_whu \\\hline
600 & 100 & 6 & 10 & 0.5 & 745    & 118     \\ \hline
\end{tabular}
\label{table1}
\end{table}

\begin{figure}[htb]
\centering
\includegraphics[width=3in]{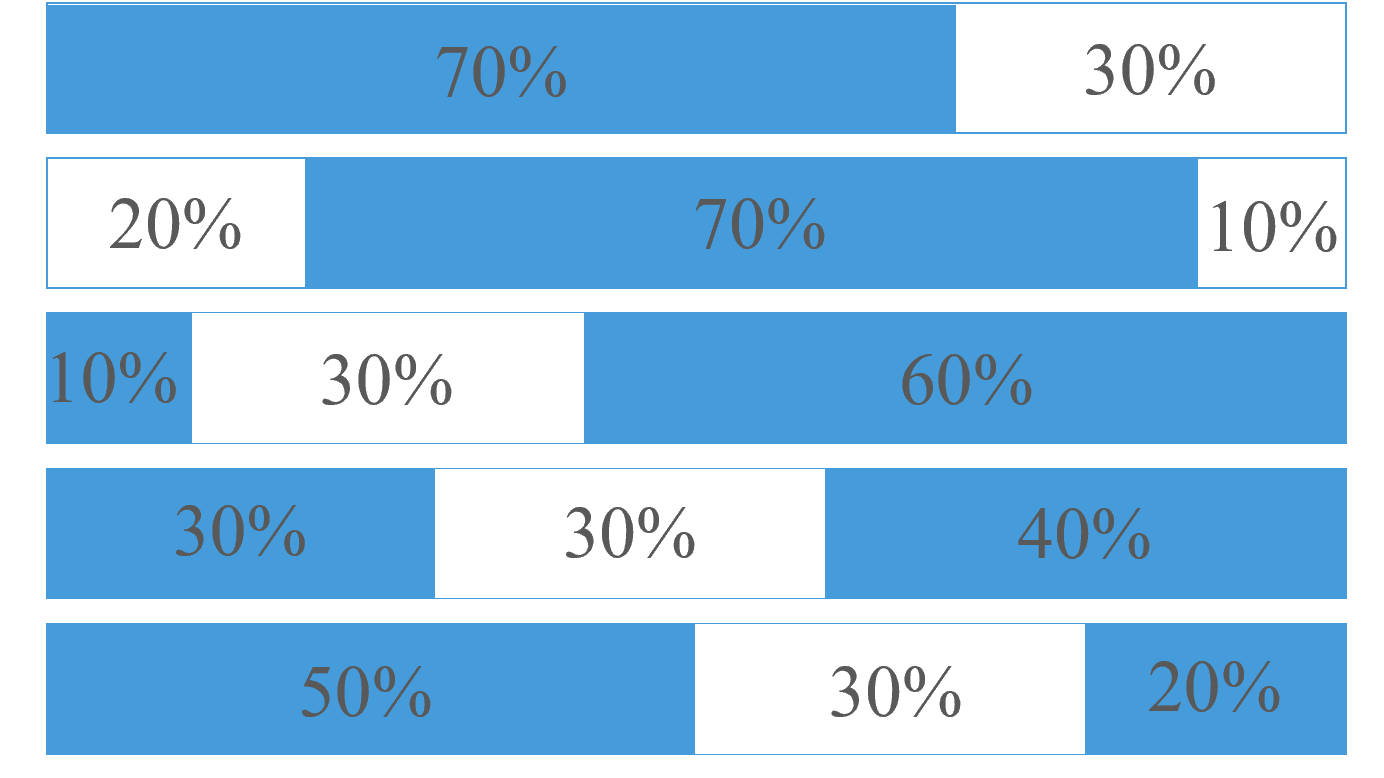}
\caption{The distribution of the cross-validation dataset.}
\label{trainset}
\end{figure}

The BLSTM network is used as a control group compared to the ABLSTM network. Then five groups of experiments are set up to make full use of the existing datasets. Each dataset is divided into a training set and a test set according to 7:3. Cross-validation is used for the datasets in each set of experiments: each set of training sets overlaps by 50$\%$, which is shown in Fig.~\ref{trainset}, where blue represents the training set and white represents the test set. We compare the ABLSTM network with three representative gait recognition networks: an LSTM model \cite{zou2020deep}, a CNN model \cite{8957063}, and an LSTM-CNN hybrid model \cite{tran2021multi}. 

\subsubsection{Result}
\begin{figure*}[htb]
\centering
\subfloat[]{\includegraphics[width=2.5in]{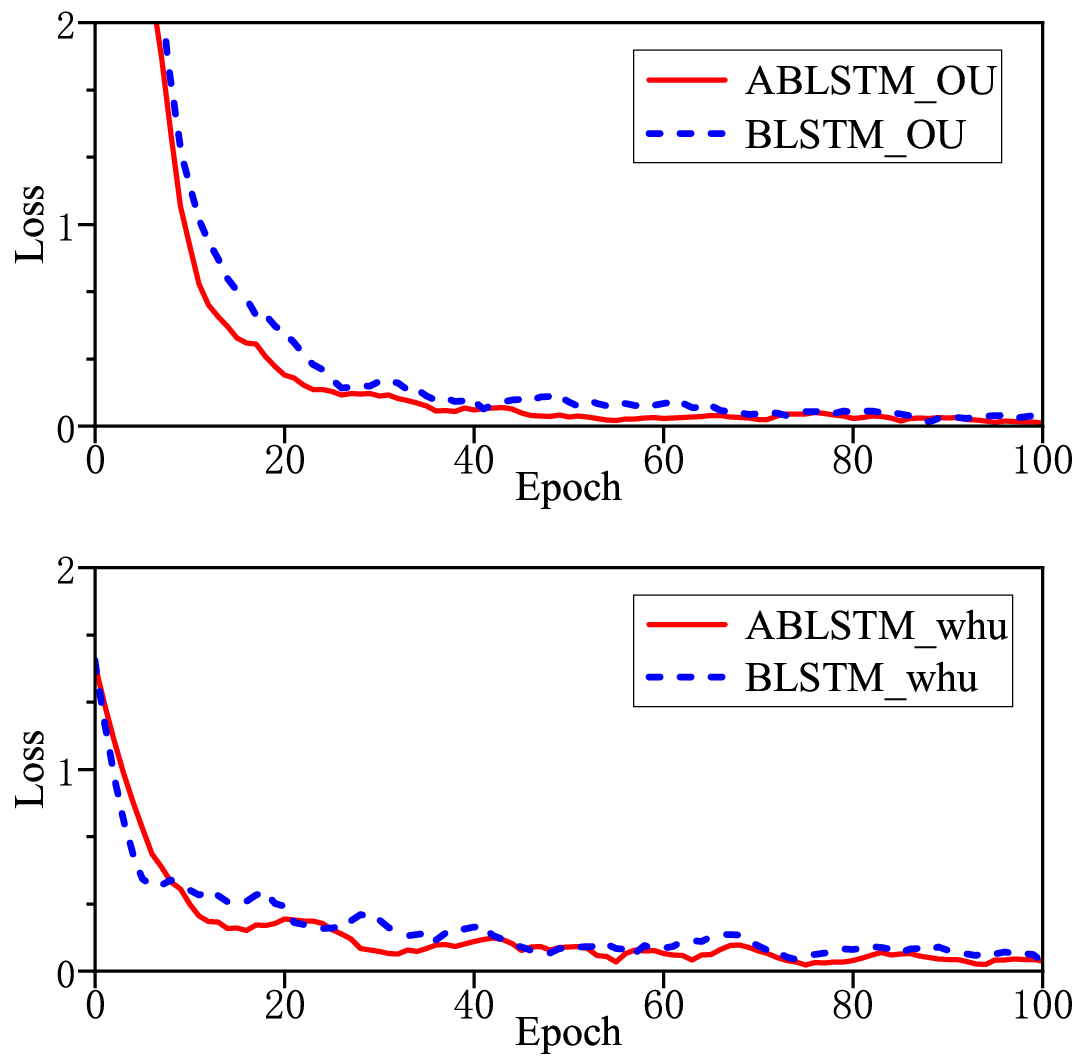}
\label{7a}}
\hfil
\subfloat[]{\includegraphics[width=2.5in]{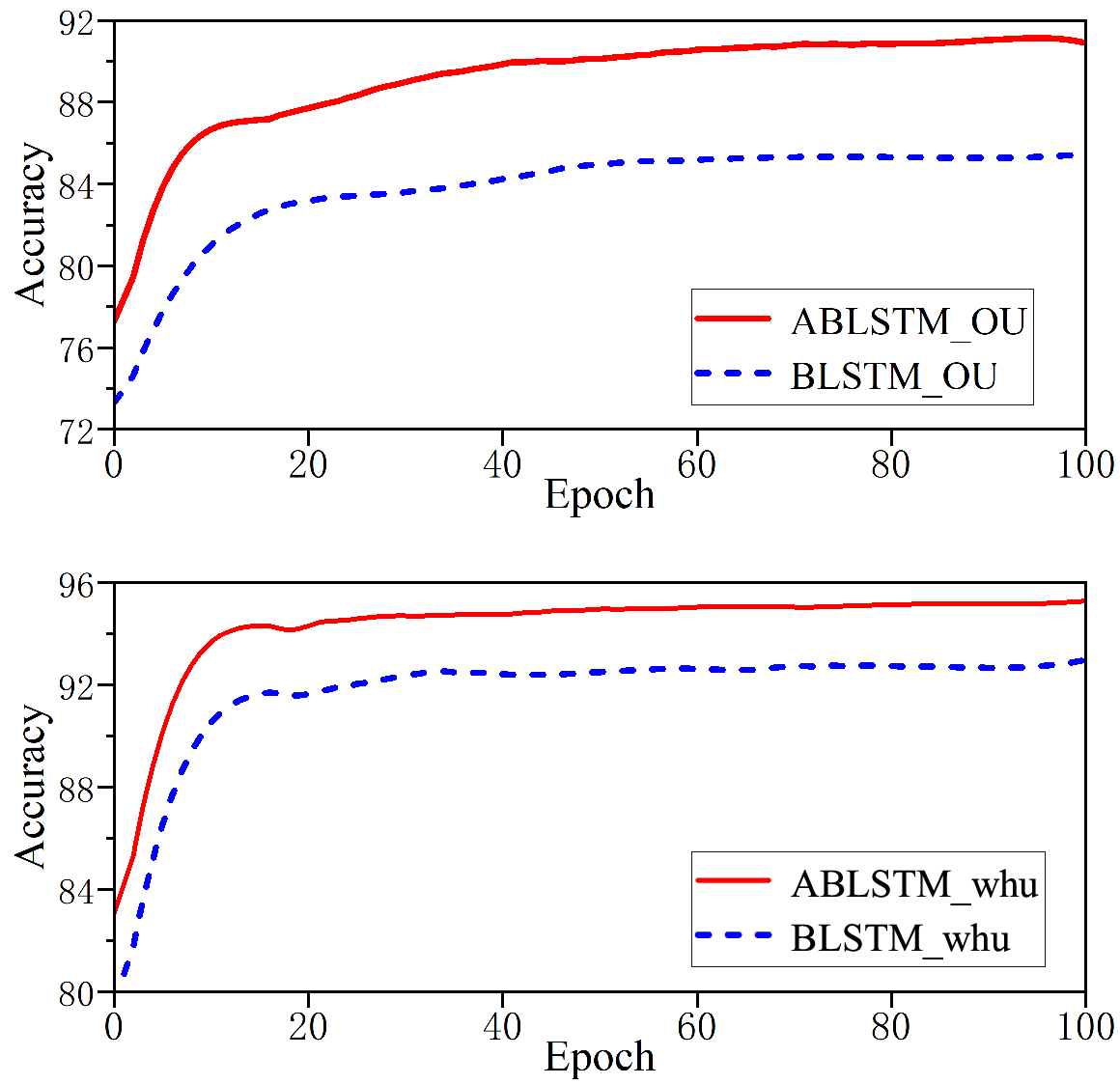}
\label{7b}}
\caption{The recognition results of gait networks. (a) The loss of BLSTM and ABLSTM. (b) The recognition accuracy of BLSTM and ABLSTM. ABLSTM\underline{~}OU and ABLSTM\underline{~}whu represent the use of the ABLSTM network on the OU-ISIR and whuGAIT datasets, respectively. BLSTM\underline{~}OU and BLSTM\underline{~}whu represent the use of the BLSTM network on the OU-ISIR and whuGAIT datasets, respectively.}
\label{t_result}
\end{figure*}

\begin{figure}[htb]
\centering
\includegraphics[width=2.5in]{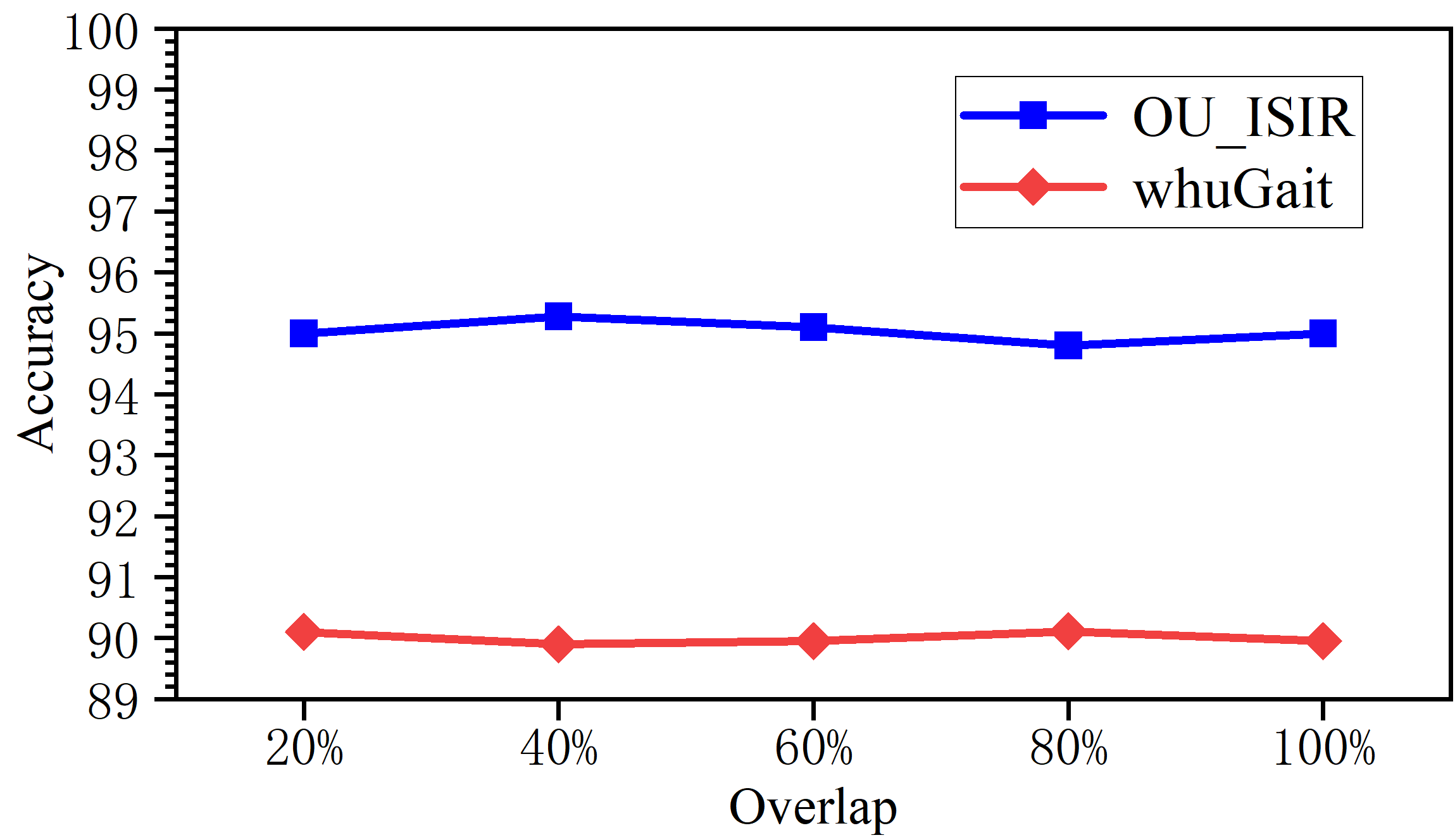}
\caption{The recognition results from dataset cross-validation.}
\label{5t_result}
\end{figure}

Figure~\ref{t_result}a shows that the ABLSTM model and the BLSTM model have almost the same loss convergence speed on the two datasets. So the ABLSTM network does not reduce the efficiency of processing gait data with full utilization of both datasets. The accuracy is expressed as the number of people correctly identified by the ABLSTM network as a percentage of the total number of people in the experiment. In Fig.~\ref{t_result}b, the ABLSTM network achieves recognition accuracy of 90.11$\%$ and 95.28$\%$ respectively. In contrast, the BLSTM achieves recognition accuracy of 86.23$\%$ and 92.11$\%$, respectively. The ABLSTM network reaches higher recognition rates compared to other gait recognition networks in Table~\ref{table2}. The recognition results show that the attention mechanism can extract gait features well and improve the recognition accuracy of the network.
Table~\ref{table2} also shows that the ABLSTM network has similar model complexity with other recognition networks. 
 In Fig.~\ref{5t_result}, corresponding to five different overlapping cross-validations of the OU-ISIR and whuGAIT datasets in Fig.~\ref{trainset}, the recognition results of ABLSTM do not fluctuate and have good robustness, demonstrating the reliable recognition accuracy of our network.
% p{1.8cm}<{\centering}p{2.5cm}<{\centering}p{2cm}<{\centering}
\begin{table}[htb] %\normalsize
\caption{The result of the recognition accuracy ($\%$) and model parameters.}
\centering
\begin{tabular}{cccc}
\hline
\textit{\textbf{Dataset}} & \textit{\textbf{Method}} & \textit{\textbf{Accuracy}} & \textit{\textbf{Model Parameters}} \\ \hline
                          & LSTM~\cite{zou2020deep}                     & 72.32           & 12M         \\
                          & CNN~\cite{8957063}                      & 82.53               &  25M    \\
\textit{OU-ISIR}          & LSTM~\cite{tran2021multi}                     & 78.92         &  12M        \\
                          & LSTM\&CNN~\cite{tran2021multi}                & 89.79         & 31M         \\
                          & \textbf{ABLSTM}         & \textbf{90.11}     &  28M      \\ \hline
                          & LSTM~\cite{zou2020deep}                     & 91.88        &  12M             \\
\textit{}                 & LSTM\&CNN~\cite{zou2020deep}              & 93.52           &  30M         \\
\textit{whuGAIT}          & CNN~\cite{8957063}                     & 93.14             &  25M     \\
\textit{}                 & LSTM\&CNN~\cite{tran2021multi}                & 94.15       &  31M             \\
\textit{}                 & \textbf{ABLSTM}          & \textbf{95.28}       &  28M    \\ \hline
\end{tabular}
\label{table2}
\end{table}

\subsection{SOT Experiment}

The SOT encryption efficiency is mainly affected by two factors: the feature's length and the number of randomly positioned parameters $a_1,\cdots, a_m$ in the SOT encryption scheme. So three different lengths templates are chosen, which are $tme480\in\mathbb{R}^{480},~tme600\in \mathbb{R}^{600},~tme960 \in \mathbb{R }^{960}$. Then 5 different numbers of randomly positioned parameters are set and the time required for different numbers of templates is recorded. The experimental results are as follows.
\begin{table}[htb]
\caption{The result of matching success rate ($\%$).}
\centering
\begin{tabular}{cccccc}
\hline
$ m$   & \bf 1      & \bf 2      & \bf 3    & \bf 4      & \bf 5      \\ \hline
$tme480$ & 88.172 & 89.458 & 87.343          & 87.646 & 86.145 \\
$tme600$ & 91.452 & 90.454 & \textbf{91.637} & 91.125 & 91.431 \\
$tme960$ & 91.034 & 90.689 & 90.678          & 90.563 & 90.727 \\ \hline
\end{tabular}
\label{table3}
\end{table}

\begin{table}[htb]
\centering
\caption{The Key Sizes}
\begin{tabular}{cccc}
\hline
 \textit{\textbf{Security Parameter}}  & \textit{\textbf{Advantage}} & \textit{\textbf{Key Size}}  & \textit{\textbf{Ciphertext Size}} \\ \hline
 \textit{$\lambda=128$}         & \textit{$2^{-266}$}       & \textit{5.68MB}   &   \textit{9.6KB}                      \\ \hline
\end{tabular}
\label{KS}
\end{table}

\begin{figure*}[htb]
\centering
\includegraphics[width=7.1in]{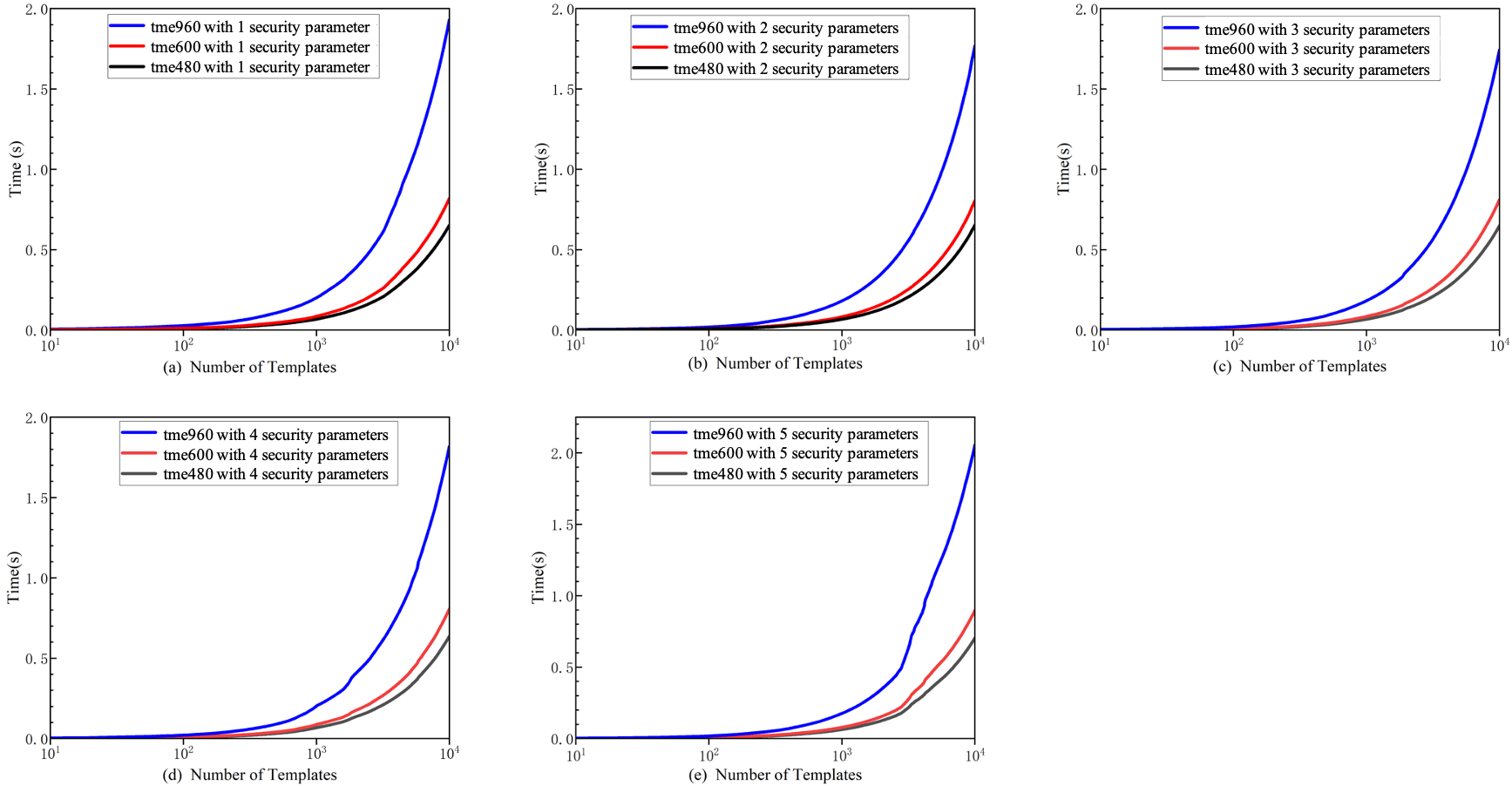}
\caption{Gait template encryption time. From (a) to (e) represent the encryption time required to add 1 to 5 randomly positioned parameters in gait templates ($tme480,~tme600,~tme960$) of 3 different lengths, respectively.}
\label{enc469}
\end{figure*}

In Fig.~\ref{enc469}, with the same number of randomly positioned parameters, the longer the template length, the greater the encryption time cost. With the same length of the templates, the time cost of using three randomly positioned parameters is the least but overall is not much different. The matching success rates of templates of various lengths corresponding to different numbers of randomly positioned parameters are shown in Table~\ref{table3}, where the columns correspond to 1 to 5 randomly positioned parameters and the rows represent templates of lengths 480, 600,  and 960. The matching success rate of the 600-length template with three randomly positioned parameters is the highest. When the security parameter is set to 128, the key size and ciphertext size are shown in Table~\ref{KS}. Based on the encryption time and recognition accuracy, adding three randomly positioned parameters of the 600-length gait template is chosen as the initial settings for our template-matching experiments. Finally, the false positive rate and true positive rate are used to evaluate the identification accuracy of our recognition network. The area under curve values on the OU-ISIR and whuGait datasets are 0.915 and 0.942, respectively. 

\begin{figure}[htb]
\centering
\includegraphics[width=7.5cm]{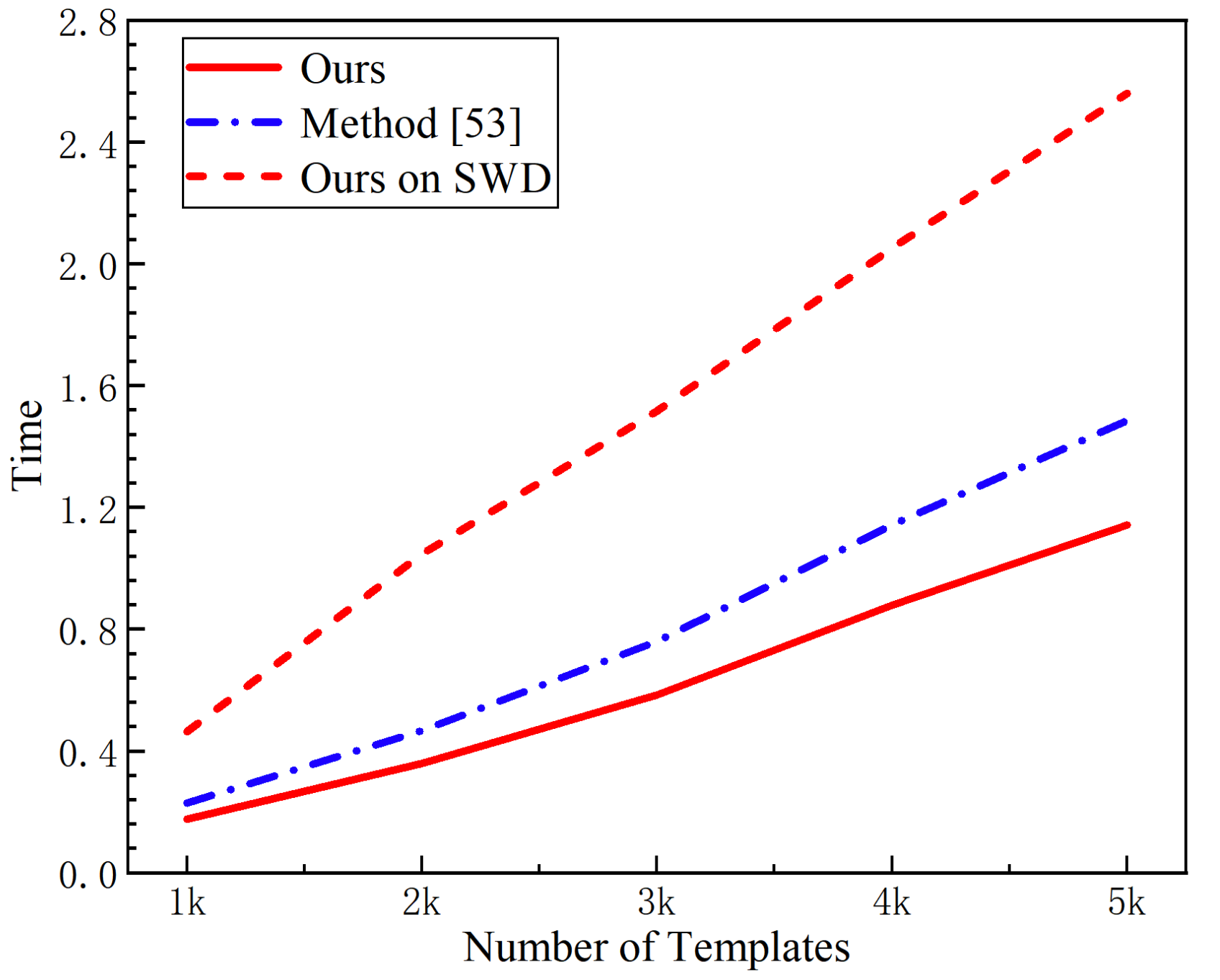}
\caption{Comparison with the Method~\cite{liu2019efficient}.}
\label{liu}
\end{figure}

Unlike the encryption method~\cite{liu2019efficient} which is applied to fingerprint templates, our encryption method is applied to gait templates with a much larger amount of data. As a comparison, both methods are experimented with using gait features of 600-length, and the results are shown in Fig.~\ref{liu}. According to our tests, our encryption method is 30$\%$ faster than the method~\cite{liu2019efficient} for different numbers of templates. 

For testing in actual scenarios, we deploy the entire encryption method to a SWD that is based on the Raspberry Pi 4b (1.5Ghz, 4 cores CPU; 4G RAM). Here the ABLSTM network uses the onnxruntime framework for deployment inference. For the same number of samples, the test time of our method on the actual device  is 2.52 times that of our method running on the computer, and 1.51 times that of the method~\cite{liu2019efficient} running on the computer, as shown in Fig.~\ref{liu}. Considering that the built-in MCU processing capability of SWD is weaker than that of pc-level processors, such a result is acceptable. Moreover, it takes only 0.04 ms to process one gait sample on the actual device.

\subsection{BBE Experiment}

The efficiency of the BBE scheme is quantified by the storage size, the amount of computation, and the encryption time. For the $n$-length feature template, the length of the feature template after preprocessing is $n+3$, the length of the master key is $(n+3)+1$ and the private key consists of only two elements. After BBE encrypts the message, the length of the ciphertext is $(n+3)+2$. For a security parameter $\lambda$ of 256, the details of key sizes are given in Table~\ref{table4}.
\begin{table*}[htb]
\centering
\caption{BBE Key Sizes}
\begin{tabular}{ccccc}
\hline
\textbf{Schemes} & \textbf{Feature Size} & \textbf{Master Key Size} & \textbf{Private Key Size} & \textbf{Ciphertext Size} \\ \hline
\textit{BBE}     & \textit{19.2KB}            & \textit{19.3KB}             & \textit{0.1KB}                & \textit{19.2KB}\\ \hline
\end{tabular}
\label{table4}
\end{table*}

The computational effort is quantified for each operation step in the BBE scheme. The quantification is based on the total number of times that the operation steps are computed on $\mathbb {G}, \mathbb {G}_T$, and $e$. The result of the quantification is shown in Table~\ref{table5}.
\begin{table*}[htb]
\centering
\caption{The Computational Effort of BBE}
\begin{tabular}{cccccc}
\hline
\textbf{Schemes} & \textbf{Master Key} & \textbf{Private Key} & \textbf{Ciphertext} & \textbf{Encryption} & \textbf{Decryption} \\ \hline
BBE              & $(n+3)|\mathbb {G}|+|\mathbb {G}_{T}|$            & $(n+4)|\mathbb {G}|+|\mathbb {G}_{T}|$           & $2|\mathbb {G}|$                & $(2n+7)|\mathbb {G}|+|\mathbb {G}_{T}|$       & $(n+3)|\mathbb {G}|+2e$ \\ \hline
\end{tabular}
\label{table5}
\end{table*}

\begin{figure}[htb]
\centering
\includegraphics[width=2.5in]{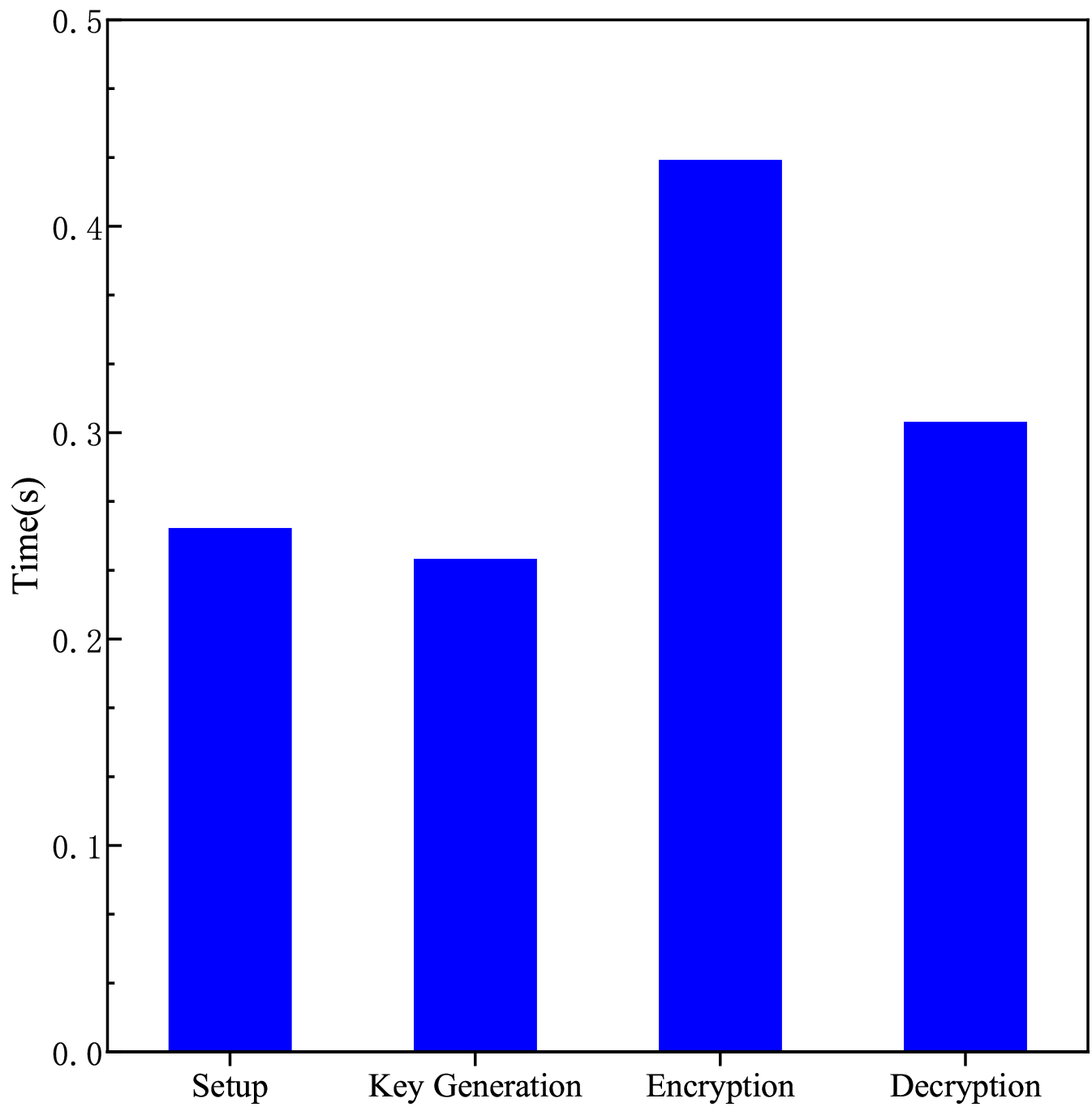}
\caption{The Time Cost of BBE.}
\label{BBET}
\end{figure}

Then the 600-length gait feature is used as the key to encrypt messages. The BBE scheme is built on the bplib 0.0.6 library. The time costs of the four algorithms of BBE are shown in Fig.~\ref{BBET}.
Finally, we provide a functional comparison of our scheme AmSoBe with previous works in Table~\ref{table9}. 
It can be seen that our scheme is the only solution that achieves all functionalities.

\begin{table}[htb]
\caption{Functional comparison. GR: Gait Recognition; FP: Feature Protection; ME: Message Encryption.}
\centering
\begin{tabular}{cccc}
\hline
  & \textbf{GR} & \textbf{FP} & \textbf{ME} \\ \hline
Tran~\cite{tran2021multi}  & $\checkmark$     &  $\times$        & $\times$       \\
Zou~\cite{zou2020deep}  & $\checkmark$     & $\times$      &  $\times$      \\
Tran~\cite{8957063}  & $\checkmark$     &  $\times$       & $\times$      \\
Liu~\cite{liu2019efficient} & $\times$     & $\checkmark$       & $\times$       \\
Ours & $\checkmark$     & $\checkmark$      & $\checkmark$       \\ 
\hline
\end{tabular}
\label{table9}
\end{table}

\section{Conclusion}
In conclusion, the ABLSTM network and the SOT scheme are proposed for SWDs. They are combined to achieve privacy-preserving gait-based identification. The ABLSTM network is applied for high-precision gait feature extraction. ABLSTM achieved 95.28$\%$ accuracy in the experiment, reducing previous error rate by 19.3$\%$. The SOT scheme is resistant to CPA to protect gait features from illegal use and is 30$\%$ faster than previous methods. In addition, a BBE scheme is proposed, which allows SWD applications to use gait features for message encryption and feature authentication. Compared to previous schemes, this BBE scheme offers better protection of the gait features.

There are several promising directions for future research. Firstly, note that multi-feature fusion has been used for identification. Hence it is an important future direction to provide relevant security protection for multi-feature identification, as an extension of the security protection for gait-feature identification. Secondly, the gait feature extraction and encryption are split into two steps in our protocol. It would be interesting to achieve end-to-end encryption in a single step, which protects the gait features even in the network training stage. A potential solution could be to combine homomorphic encryption with the neural network.

\section*{Note added}
After completion of this work, we became aware of an independent work \cite{hasan2022gait} which achieves similar gait recognition accuracy. They used a different gait network, and did not consider the storage security of the gait features and secure message interactions between SWDs and other devices.

\bibliographystyle{IEEEtran}        % Include this if you use bibtex 
\bibliography{IEEEabrv,ref}

\end{document}